\documentstyle[tighten,preprint,eqsecnum,aps,prd,floats]{revtex}
\input{epsf}
\input{psfig}

\def\no{\nonumber}

\def\e{\eta}

\def\be{\begin{equation}}
\def\bea{\begin{eqnarray}}
\def\eea{\end{eqnarray}}
\def\ee{\end{equation}}
\def\bi{\begin{itemize}}
\def\ei{\end{itemize}}

\begin{document}
\draft

\preprint{\vbox{\baselineskip=12pt
\rightline{}
gr-qc/0111064}}
\title{On the observational determination of squeezing in relic 
gravitational waves and primordial density perturbations} 
\author{Sukanta Bose\footnote{Permanent address:
Department of Physics, Washington State University, 1245 Webster, Pullman,
WA 99164-2814, U.S.A.; electronic address:
{\em sukanta@wsu.edu}}(1,2) and 
L. P. Grishchuk\footnote{Electronic address:
{\em grishchuk@astro.cf.ac.uk}}(1,3)}
\address{{\rm (1)} Department of Physics and Astronomy, P. O. 
Box 913, Cardiff University, CF24 3YB, United Kingdom}
\address{{\rm (2)} Max Planck Institut f\"{u}r Gravitationsphysik,
Albert Einstein Institut, Am M\"{u}hlenberg 1,\\ Golm, D-14476, Germany}
\address{{\rm (3)} Sternberg Astronomical
Institute, Moscow University, Moscow 119899, Russia}
\date{November 2001}

\begin{abstract}
We develop the theory in which relic gravitational waves and primordial density 
perturbations are generated by strong variable gravitational field of the 
early Universe. The generating mechanism is superadiabatic (parametric) 
amplification of the zero-point quantum oscillations. The generated fields 
have specific statistical properties of squeezed vacuum quantum states. 
Macroscopically, squeezing manifests itself in a non-stationary character of 
variances and correlation functions of the fields, periodic structures of the 
metric power spectra, and, as a consequence, in oscillatory behavior of the 
higher order multipoles $C_l$ of the cosmic microwave background anisotropy. We
start with the gravitational wave background and then apply the theory to 
primordial density perturbations. We derive an analytical formula for the 
positions of peaks and dips in the angular power spectrum $l(l+1)C_l$ as a 
function of $l$. This formula shows that the values of $l$ at the peak 
positions are ordered in the proportion $1:3:5:...$, whereas at the dips they 
are ordered as $1:2:3:...$. We compare the derived positions with the actually 
observed features, and find them to be in reasonably good agreement. It appears
that the observed structure is better described by our analytical formula based
on the (squeezed) metric perturbations associated with the primordial density 
perturbations, rather than by the acoustic peaks reflecting the existence of 
plasma sound waves at the last scattering surface. We formulate a forecast for 
other features in the angular power spectrum, that may be detected by the 
advanced observational missions, such as MAP and PLANCK. We tentatively 
conclude that the observed structure is a macroscopic manifestation of 
squeezing in the primordial metric perturbations.     
\end{abstract}
\pacs{98.80.Cq, 98.70.Vc, 98.80.Es, 42.50.Dv, 04.30.-w}

\maketitle

\narrowtext

\vfil
\pagebreak

\section{Introduction}
\label{sec:intro}

A direct search for relic gravitational waves is one of the important goals
of the forthcoming gravitational wave observations (for reviews, see Refs. 
\cite{thorne}, \cite {schutz}, \cite{ufn}). Relic gravitational waves are 
inevitably generated by strong variable gravitational field of the 
very early Universe through the mechanism of superadiabatic (parametric) 
amplification of the zero-point quantum oscillations \cite {LPG1}.
The word ``superadiabatic" emphasizes the fact that this effect takes 
place over and above whatever effects occur during very 
slow (adiabatic) changes. That is, we are interested in the increase 
of occupation numbers, rather than
in the gradual shift of energy levels.  The word ``parametric" emphasizes
the underlying mathematical structure of the wave equations. It is a 
sufficiently quick change of a parameter of the oscillator, namely,
variation of its properly defined frequency, that is responsible for 
the considerable increase of energy of that oscillator.      

Apparently, Schr\"{o}dinger \cite{sch} was the first to notice the
``alarming phenomenon" in an expanding universe. Specifically, Schr\"{o}dinger
discusses the ``mutual 
adulteration of positive and negative frequency terms in the course of 
time". The frequency mixing means that a traveling wave 
can be amplified, with the simultaneous appearance of a ``reflected" wave, 
i.e., a wave traveling in the opposite direction. After Schr\"{o}dinger, out 
of unawareness of his work, this effect has been rediscovered several
times. Schr\"{o}dinger speaks about the mutual adulteration of 
electromagnetic waves, which would mean the generation of photons.  
We now know that the coupling of the electromagnetic field to gravity is
such that the generation of photons is impossible, so that the alarming
phenomenon does not take place. A detailed study of the Schr\"{o}dinger 
paper shows that, in fact, he was operating with a variant of scalar
electrodynamics, that is, with a scalar wave equation in an expanding
universe model (for a discussion of this point, see Ref. \cite{gr2}). Then,
indeed, the coupling of a scalar field to gravity can be chosen in such 
a way (minimal coupling), that the generation of scalar particles becomes 
possible. Parker \cite{Parker}  
undertook a systematic study of the quantized version
of the scalar wave equation in FLRW (Friedman-Lemaitre-Robertson-Walker)
cosmologies. For a summary of the subject, see Ref. 
\cite{grib}, \cite{davies}. 
As for gravitational waves, there is no ambiguity in their 
coupling to gravity because the coupling follows directly from the 
Einstein equations. It was shown that the gravitational wave equation 
for each of the two polarization components is exactly the same
as the equation for the minimally-coupled massless scalar field \cite{LPG1}. 
The early studies were concerned with free test fields superimposed
on a given space-time, whereas we are interested in fields arising in
the context of perturbed Einstein equations (cosmological perturbations).
This distinction is especially important for the issue of quantum 
normalization of the fields.

Already at this elementary level of discussion, one can make  
an important observation that will play a crucial role in our study below.
If a classical traveling wave, of any physical nature, is going 
to be strongly amplified,
the resulting wave-field will form an almost standing wave. A traveling wave
can never convert itself into a strict standing wave, because of the 
conservation of linear momentum. But the final amplitudes of the 
amplified left-moving and right-moving waves will be large and almost 
equal, so they interfere to form a practically standing wave.

The amplification process is linear, and the final amplitude of a classical
wave is proportional to the initial amplitude. If the amplitude of
a classical oscillator is zero initially, the oscillator will not get
excited by the parametric influence. However, a quantum oscillator in its
vacuum state does possess tiny ``zero-point" quantum oscillations. 
One can think of these vacuum oscillations as the ones
that are being amplified. The generation of relic 
gravitational waves (as well as the generation of other cosmological 
perturbations, discussed 
below) is a genuine quantum-gravity process, in the sense that the final
result inherently contains all the fundamental constants $\hbar$, G, and c.
The gravitational energy-momentum tensor contains G and c, while the Planck
constant $\hbar$ enters through the requirement of having initial
energy $\frac{1}{2}\hbar \omega$ per mode of the perturbation
field. The fundamental constants
naturally combine in the Planck length $l_{Pl} = (G \hbar/c^3)^{1/2}$ or
the Planck mass $m_{Pl} = (\hbar c/G)^{1/2}$, but $l_{Pl}$ or $m_{Pl}$
must stay in the numerator of the final expression, not in its denominator, 
so that the final result vanishes if $\hbar$ is formally sent to zero.     

Gravitational field of a FLRW universe is given by the metric 
\begin{equation}\label{cosmet}
{\rm d}s^2 = - c^2{\rm d}t^2 + a^2(t) g_{ij} {\rm d}x^i{\rm d}x^j = 
a^2({\eta})[-{\rm d}\eta^2 + g_{ij} {\rm d}x^i{\rm d}x^j] \ \ ,
\end{equation}
where the scale factor $a(t)$ (or $a(\eta)$) is driven by
matter distribution with some effective (in general, time-dependent)
equation of state. The scale factor has the dimensionality of length,
while $\eta$ and $x^i$ are dimensionless. 
Without restricting in any way the physical content of the problem, 
one can write the perturbed gravitational field of a FLRW 
universe (for simplicity, spatially flat) as   
\begin{equation}
\label{cosmetgw}
{\rm d}s^2 = a^2({\eta})[-{\rm d}\eta^2 + (\delta_{ij} + h_{ij})
{\rm d}x^i{\rm d}x^j]  ,
\end{equation} 

\begin{eqnarray}
\label{hij}
h_{ij} (\eta ,{\bf x})
= \frac{\cal C}{(2\pi )^{3/2}} \int_{-\infty}^\infty d^3{\bf n}
  \sum_{s=1, 2}~{\stackrel{s}{p}}_{ij} ({\bf n})
   \frac{1}{\sqrt{2n}}
\left[ {\stackrel{s}{h}}_n (\eta ) e^{i{\bf n}\cdot {\bf x}}~
                 {\stackrel{s}{c}}_{\bf n}
                +{\stackrel{s}{h}}_n^{\ast}(\eta) e^{-i{\bf n}\cdot {\bf x}}~
                 {\stackrel{s}{c}}_{\bf n}^{\dag}  \right]. 
\end{eqnarray}

The functions $h_{ij} (\eta ,{\bf x})$ have been expanded over spatial 
Fourier harmonics $e^{\pm i{\bf n}\cdot {\bf x}}$, where
${\bf n}$ is a constant (time-independent) wave vector. 
The wave number, $n$, is related to ${\bf n}$ by 
$n = (\delta_{ij}n^in^j)^{1/2}$. The wave number $n$ defines the wavelength
measured in units of laboratory standards
(so to say, in centimeters) by $\lambda = 2 \pi a/n$.  
Using the Fourier expansion, we are able to reduce the perturbed dynamical
problem to the evolution of mode functions 
${\stackrel{s}{h}}_n (\eta )$ for each mode ${\bf n}$. 
Two polarization tensors ${\stackrel{s}{p}}_{ij}({\bf n}),~s=1,2$ have
different forms depending on whether they represent gravitational waves,
rotational perturbations, or density perturbations. If $( {\bf n}/n, {\bf l},
{\bf m})$ are three unit and mutually orthogonal (spatial) vectors, then we have 
for gravitational waves,
\[ 
{\stackrel{1}{p}}_{ij} = l_i l_j - m_j m_i,~~   
{\stackrel{2}{p}}_{ij} = l_i m_j + l_j m_i,~~   
{\stackrel{s}{p}}_{ij}\delta^{ij} = 0, ~~
{\stackrel{s}{p}}_{ij}n^j = 0,
\]
for rotational perturbations,
\[
{\stackrel{1}{p}}_{ij} =\frac{1}{n} (l_i n_j + l_j n_i),~~   
{\stackrel{2}{p}}_{ij} =\frac{1}{n} (m_i n_j + m_j n_i),~~   
{\stackrel{s}{p}}_{ij}\delta^{ij} = 0, ~~
{\stackrel{s}{p}}_{ij}n^in^j = 0,
\]
and for density perturbations,
\[
{\stackrel{1}{p}}_{ij} = \sqrt{\frac{2}{3}} \delta_{ij},~~
{\stackrel{2}{p}}_{ij} = -\sqrt{3}\frac{n_i n_j}{n^2} +
\frac{1}{\sqrt{3}} \delta_{ij}.
\]
In all three cases, ${\stackrel{s}{p}}_{ij}({\bf n})$ obey 
\[
 {\stackrel{s'}{p}}_{ij} ({\bf n})
 {\stackrel{s}{p}}~^{ij} ({\bf n}) = 2\delta_{ss'}, ~~
 {\stackrel{s}{p}}_{ij}(-{\bf n}) = {\stackrel{s}{p}}_{ij}({\bf n}). 
\]

In general relativity, rotational and density perturbations can only
exist if they are supported by the corresponding perturbations of 
matter. Their propagation speeds depend on the properties of matter and
can range from zero to the speed of light, $c$. For instance, the
propagation speed of density perturbations in the radiation-dominated fluid
is $c/\sqrt 3$; and it was very close to $c$ if the very early universe
was driven by a scalar field \cite{gri94}. However, in alternative 
theories of gravity, solutions with the polarization structure of rotational
and density perturbations can exist even in the absence
of matter fields, in which case the metric perturbations
represent gravitational waves with new polarization states, in addition to
the usual gravity-wave polarization states of general relativity
\cite{will}. If one concentrates on metric perturbations alone, temporarily 
leaving aside the accompanying perturbations of matter variables, then 
all three types of cosmological perturbations in general relativity can be 
thought of
as gravitational waves, even though some of them have unusual polarization 
states and unusual propagation speeds. There is no wonder that the dynamical 
equations for cosmological rotational and density perturbations are 
similar to, and sometimes exactly the same as, equations for cosmological
gravitational waves. The common ``master equation",
whose solutions allow one to derive all the metric components along with
all the matter perturbations (when they are present), has the 
universal form \cite{gri94}:
\begin{equation}
\label{master}
f^{\prime\prime} + f \left[n^2 \frac{c_l^2}{c^2} - W(\eta) \right] = 0 \ \ ,   
\end{equation}
where $\prime := {\rm d}/{\rm d} \eta$, $c_l$ is a function of $\eta$ and is
interpreted as the propagation speed of the perturbation, 
and $W(\eta)$ is a function of $a(\eta)$ and its derivatives. For
density perturbations in a perfect fluid with the fixed equation of
state $p = w \epsilon$, $c_l$ is a constant. In the case
of gravitational waves, $c_l^2/c^2 = 1$ and $W(\eta) = a^{\prime\prime}/a$ 
\cite{LPG1}. One can view Eq. (\ref{master}) as the equation of an oscillator 
with variable frequency (the term in square brackets), or as the 
Schr\"{o}dinger equation of a particle moving in the presence of a potential 
barrier $W(\eta)$ (while
remembering that $\eta$ is a time coordinate rather than a spatial coordinate).
In what follows, we will be discussing gravitational waves and 
density perturbations.       
 
For a classical gravitational field, the quantities  
${\stackrel{s}{c}}_{\bf n},~{\stackrel{s}{c}}_{\bf n}^{\dag}$ in
Eq. (\ref{hij}) are arbitrary complex-conjugate numbers. 
The constant $\cal C$ can be incorporated into them. In the
quantized version, the quantities  
${\stackrel{s}{c}}_{\bf n},~{\stackrel{s}{c}}_{\bf n}^{\dag}$ are
annihilation and creation operators satisfying the conditions
\be\label{comm}
[{\stackrel{s'}{c}}_{\bf n},~{\stackrel{s}{c}}_{{\bf m}}^{\dag}]=
\delta_{s's}\delta^3({\bf n}-{\bf m})\>, \quad 
{\stackrel{s}{c}}_{\bf n}|0\rangle =0 \ \ , 
\ee
where $|0\rangle$ (for each ${\bf n}$ and $s$) is the fixed 
initial vacuum state defined at some $\eta_0$ in the very distant past, long
before the superadiabatic regime for the given mode has started. In that 
early era, the mode functions ${\stackrel{s}{h}}_n (\eta )$ 
behaved as $\propto e^{-i n \eta}$, so that each mode $\bf n$ represented a
strict traveling wave propagating in the direction of $\bf n$. 
The normalization constant $\cal C$ is $\sqrt{16 \pi} l_{Pl}$ for
gravitational waves, and $\sqrt{24 \pi} l_{Pl}$ for density perturbations.  
 
A detailed study shows \cite{GS90} that the quantum-mechanical Schr\"{o}dinger 
evolution brings the initial vacuum state of cosmological perturbations 
into the final multi-quantum state known as squeezed vacuum state. It is the 
variance of phase that is being strongly diminished (squeezed), while the mean 
number of quanta and its variance are being strongly increased. A squeezed 
vacuum state is conveniently characterized by the squeeze parameter $r$. The 
squeeze parameter grows from $r=0$ in the vacuum state up to $r \gg 1$ by the 
end of the amplifying superadiabatic regime. The mean number of quanta in a 
2-mode squeezed vacuum state is $ \langle N \rangle = 2 \sinh^2 r$. Squeezed 
vacuum states possess specific statistical
properties. In particular, the generated field, viewed as a random field,
obeys the statistics of a Gaussian non-stationary process. The 
non-stationarity means that the variance of the field is an explicit 
oscillatory function of time, and the two-time correlation function 
depends on individual moments of time, not only on the time difference.    
The calculation of quantum-mechanical expectation values and correlation 
functions provides the link between quantum mechanics and macroscopic 
physics. 

Using the representation (\ref{hij}) and definitions above, one finds the 
variance of metric perturbations: 
\begin{equation} 
\label{hmvar}
\langle 0| h_{ij}(\e,{\bf x}) h^{ij}(\e,{\bf x})|0\rangle 
= \frac{{\cal C}^2}{2\pi^2} \int_{0}^{\infty} n^2\sum_{s=1,2}
|{\stackrel{s}{h}}_n(\e)|^2 \frac{{\rm d}n}{n}.
\end{equation}
The quantity 
\begin{equation}
\label{power}
h^2(n, \eta) =  
\frac{{\cal C}^2} {2\pi^2} n^2\sum_{s=1,2} |{\stackrel{s}{h}}_n(\e)|^2 
\end{equation}
gives the mean-square value of the gravitational field perturbations in
a logarithmic interval of $n$ and is called the (dimensionless) power 
spectrum.  
In the case of gravitational waves, it is relatively easy to evolve the 
mode functions up to the present era, and to find that
\begin{equation}
\label{structure} 
h^2(n, \eta) \propto \sin^2[n(\eta - \eta_e)],
\end{equation}
where $\eta_e$ is a constant discussed below. The explicit time-dependence
of the power spectrum is a consequence of squeezing and can be also
viewed as a reflection of the standing-wave pattern of the generated
field. For every fixed moment of time (for instance, today) the power 
spectrum contains many maxima and zeros at certain wave numbers, 
even though the spectrum
was perfectly smooth before amplification, i.e., when the mode functions 
${\stackrel{s}{h}}_n (\eta )$ behaved as $\propto e^{-i n \eta}$.
As soon as the amplifying 
process takes place, the increase of the mean number of quanta, squeezing,
non-stationarity, formation of standing wave pattern and oscillatory
features in the power spectrum, - are all the different facets of the same
phenomenon. 

The relative spacing of zeros is very dense at
laboratory scales (large $n$'s), but is quite sparse at cosmological
scales (small $n$'s). Specifically, the spectrum contains about $10^{20}$
zeros in the interval from 100 Hz to 200 Hz, but only a dozen of zeros
in the interval from 1000 Mpc to 2000 Mpc. The oscillatory 
time-dependence (\ref{structure}) is known in advance, and this information  
would certainly help, in a very narrow-band gravitational wave detector, 
to find the signal against the instrumental noise, and to provide evidence 
for the 
primordial origin of the detected gravitational wave background \cite{ufn}. 
However, in a broad-band detector,
there are too many zeros together, and the non-stationary process is 
practically indistinguishable from the stationary process of the same
power density. In a recent
paper, Allen, Flanagan, and Papa \cite{AFP} agree that the non-stationarity
in the relic background is present, but they argue that this feature can 
hardly give any advantage in practice. Without presently being able to offer a 
realistic scheme of exploiting the non-stationarity at small scales,
we shift our attention to cosmological scales, where the spacing of zeros is   
sparse. The natural place to look for the consequences of squeezing is
the distribution of multipoles $C_\ell$ of the cosmic microwave
background (CMB) anisotropies \cite{gristat}.  

Relic gravitational waves should be important at lower multipoles 
$\ell$ of the CMB anisotropies, but they are not expected to give significant 
contribution at $\ell \sim 200$ and higher. It is the density 
perturbations that are expected to be primarily responsible for the behavior
of the $C_\ell$'s in the latter
range. However, we study in great detail the simpler case of gravitational
waves in order to resolve a number of principal issues and to get a guidance 
for the analysis of the technically more difficult case of density 
perturbations. The phenomenon of squeezing is universal, and if the 
primordial density perturbations have a quantum-mechanical origin (which, 
we believe, is likely to be true), then many features must be 
common with the case
of relic gravitational waves. Some differences arise at the late stages
of evolution (in particular, they explain why the gravitational wave ($g.w.$) 
contribution is
subdominant at $\ell \sim 200$ and higher), but we take them into account. 
Developing the conjecture
of Ref. \cite{dg}, we argue that it is the modulated structure of the
metric power spectrum (which is caused, on fundamental level, by squeezing) 
that is responsible for the downturn of the rising 
function $\ell (\ell + 1) C_{\ell}$ at the peak, and for 
the appearance of subsequent peaks and dips, a few of which have been 
recently observed \cite{peaks}. On the ground of our simple analytical 
treatment we make a 
forecast for the positions of further peaks and dips that may be observed by
future missions, such as MAP and PLANCK. In general, our forecast 
agrees with that of Ref. \cite{deb}, made on the grounds of numerical 
codes, but discrepancies become significant 
somewhere around the 4-th expected peak.  

The structure of the paper and its conclusions are as follows.  
We start, in Sec. \ref{sec:regw}, with the gravitational wave solutions in the 
present universe, that is, at the matter-dominated stage. We consider the
general solution for the time dependent mode functions
${\stackrel{s}{h}}_n (\eta )$, regardless of whether a given
particular solution is likely to emerge from the very early universe or not.
In general, a given mode ${\bf n}$ is neither a traveling wave nor a
standing wave. We formulate conditions under which a given mode is a
strict traveling wave or a strict standing wave. These conditions are
constraints on the (Fourier) coefficients $A_n, B_n$ in front of two 
linearly independent solutions for the mode functions. 
Then, we explore the issue of stationary versus non-stationary variance.
We demonstrate that oscillations in the power spectrum are most pronounced
when the modes are standing waves, and they disappear when the modes are
traveling waves. Thus, the often made (incorrect) statement that the 
relic gravitational wave background should be stationary is equivalent 
to the assumption that it is being formed by traveling waves. We return to 
this issue later on (in Sec. \ref{sec:squeeze}) and show that, whether the 
non-stationarity on small scales is measurable or not, the very assumption of 
the stationary gravitational wave background is in conflict with some other 
cosmological considerations. In Sec. \ref{sec:model}, we present a ``physical" 
model for the gravitational wave background. We call it ``physical" because
we evolve the field through all the three relevant stages of cosmological 
evolution -- initial, radiation-dominated, and matter-dominated. In this
way we distinguish the ``physical" model from the ``alternative" model, 
which postulates that the
gravitational wave background is stationary, irrespective of its 
physical origin. The waves generated in the physical model are
squeezed (standing). We demonstrate that the evolution of standing waves
through the effective barrier $a^{\prime \prime}/a$ at the matter-dominated
stage results in the appearance of an oscillatory behavior of the Fourier
coefficients $A_n, B_n$ as functions of the wave number $n$. We later
show (in Sec. \ref{sec:anisotropies}) that this oscillatory behavior of 
$A_n, B_n$ is the origin of oscillations 
in the multipole moment distribution $C_{\ell}$ as a function of $\ell$. 

Section \ref{sec:squeeze} compares the physical and alternative gravitational 
wave backgrounds. We introduce the notion of a fair comparison, which requires 
that today's band-powers of the two backgrounds be equal at all scales.
By evolving the corresponding solutions backwards in time, we show that
the alternative background would have had too much power in long waves
at the era of last scattering of the CMB radiation. In terms 
of ``growing" and ``decaying" solutions, this means that the amplitude
of the decaying solution in the alternative background becomes 
dangerously large when one returns deeper and
deeper in the past. The further evolution backwards in time would have 
destroyed our sacred belief that the Universe was homogeneous and isotropic
(up to small perturbations) at the time of the primordial nucleosynthesis 
and its past. Most
importantly for our study, we demonstrate that the alternative background
does not produce oscillations in the $C_{\ell}$ multipoles. This shows that
the squeezing is observationally distinguishable, even if, at the present level
of observational capabilities, it is better to search on large scales rather
than on small scales. 

Section \ref{sec:anisotropies} presents the results of numerical calculations
of the $C_{\ell}$'s caused by relic gravitational waves. We show that 
our analytical
formula for the positions of peaks and dips is in a fairly good agreement 
with numerical calculations. This analysis demonstrates that,
at least in the case of gravitational waves, the $C_{\ell}$ oscillations 
are produced by modulations in the power spectrum of metric 
perturbations, and not by acoustic waves at the last scattering 
surface, simply because there are no matter perturbations at all. 

In Section \ref{sec:dp}, we 
turn to the primordial density perturbations. The evolution of density
perturbations through the initial and radiation-dominated stages is
almost identical to the evolution of gravitational waves. We 
show that the Fourier coefficients $A_n, B_n$ of metric perturbations 
associated with density perturbations develop 
a periodic structure, as functions of $n$, in the course of 
transition from the radiation-dominated phase to the matter-dominated
phase. In a manner similar to the gravitational wave case, we derive the
expected positions of peaks and dips in the $C_{\ell}$ distribution. 
Because of the damping, features beyond the second peak may not be easily 
discernible \cite{wein1}, \cite{wein2}, \cite{wein3}, but they
seem to be less likely to be washed out if they are produced by modulations 
in the metric power spectrum rather than by modulations in the plasma matter 
power spectrum. We show that the peak positions should obey the rule
1:3:5:7..., whereas the dip positions should be ordered as 1:2:3:4.... We
demonstrate that the observed positions agree better with our analytical
formula than with the concept of ``acoustic peaks". The 
detection of late features may be especially
interesting as our analytical forecast is somewhat different from 
what follows from conventional numerical codes. We 
tentatively conclude that the observed structures in the
angular power spectrum $\ell(\ell +1)C_{\ell}$ 
are macroscopic manifestations of squeezing in gravitational 
field perturbations.

\section{Properties of gravitational wave solutions}
\label{sec:regw}

The perturbed Einstein equations for gravitational waves can be reduced
to the ``master equation"
\begin{equation}\label{fieldeq}
{\stackrel{s}\mu}_{n}^{\prime\prime} + {\stackrel{s}\mu}_{n} \left[n^2 - 
\frac{a^{\prime\prime}}{a}\right] = 0 \  ,   
\end{equation}
where the functions ${\stackrel{s}\mu}_n(\e)$ are related to the mode 
functions ${\stackrel{s}{h}}_n(\eta )$ by 
\be \label{mudef}
{\stackrel{s}{\mu}}_n (\e) \equiv  a(\e) {\stackrel{s}{h}}_n (\eta ) .
\ee
We suppress the polarization index $s$ when it causes no ambiguity.

The scale factor $a(\eta)$ at the matter-dominated stage, 
governed by whatever matter with the effective equation of state $p=0$, 
behaves as $a(\eta) \propto \eta^2$. It is convenient to write $a(\eta)$ 
in the explicit form
\be\label{scale-matter}
a(\e) = 2l_H (\e-\e_m)^2 \  , 
\ee
where $l_H$ is the Hubble radius today ($l_H = c/H_0$, where $H_0$ is the 
present value of the Hubble parameter) and $\e_m$ is a constant explained
below. The moment of time ``today" (in cosmological sense) is labeled 
by $\eta=\eta_R$ (the subscript $R$ denoting ``reception'').  
It is convenient to choose 
\begin{equation} \label{etaR}
\eta_R - \eta_m =1. 
\end{equation}
With this convention,
$a(\eta_R) = 2 l_H$, and the wave, of any physical nature, whose 
wavelength $\lambda$ today is equal to today's
Hubble radius, carries the constant wavenumber $n_H = 4 \pi$. Longer waves 
have smaller $n$'s and shorter waves have larger $n$'s, 
according to the relationship $n = 4 \pi l_H/ \lambda$. For example,
the ground-based gravitational wave detectors are most sensitive to 
frequencies around 30 Hz - 3000 Hz. The corresponding wavelengths have 
wavenumbers $n$ somewhere in the interval $10^{20} - 10^{22}$.  

For the scale factor (\ref{scale-matter}), 
Eq. (\ref{fieldeq}) is easily solved to yield
\bea\label{mumgen}
\mu_{n} &=& \sqrt{y} \left[ A_n J_{3/2}(y) -iB_n J_{-3/2}(y)\right] \no\\
&\equiv& \frac{\sqrt{y}}{2}\left[ \left(A_n -B_n \right) H^{(1)}_{3/2}(y) 
               + \left(A_n +B_n \right) H^{(2)}_{3/2}(y)\right] \ \ ,
\eea
where 
\be
y \equiv n(\e-\e_m) \ \ , 
\ee
and $J_{\pm 3/2}(y)$ and $H^{(1,2)}_{3/2}(y)$ are Bessel and Hankel 
functions, respectively \cite{AS}. We will also be using spherical Bessel 
functions $j_i (y) = \sqrt{\pi/2y} J_{i+1/2}(y)$.  
The (Fourier) coefficients $A_n$ and $B_n$ 
are, so far, arbitrary complex numbers.
In the limit $y\to \infty$,
\be\label{asym} 
H^{(1)}_{3/2}(y) \sim -\sqrt{2\over \pi y}\> e^{iy},~~ 
H^{(2)}_{3/2}(y) \sim -\sqrt{2\over \pi y}\> e^{-iy}~~{\rm and}~~
j_1(y) \sim -{1 \over y} \cos y,~~j_{-2}(y) \sim -{1 \over y} \sin y . 
\ee
Thus, at late times, $\mu_{n}(\eta)$ is a combination of sine and cosine 
functions of time:
\bea\label{mumgenLate}
\mu_{n}(\eta) &=& -{1 \over \sqrt{2\pi}}\left[ \left(A_n -B_n \right) e^{iy}
               + \left(A_n +B_n \right)  e^{-iy} \right]  \no\\
 &=& -\sqrt{2 \over \pi} \left( A_n \cos y -i B_n \sin y \right) . 
\eea

\subsection{Standing versus traveling waves}
\label{subsec:stand}

We now consider a {\em classical} field (\ref{hij}). The ${\bf n}$-th
mode of the field is given by the real function
\bea\label{nth}
h_{\bf n} (\e,{\bf x}) \equiv  h_n (\eta ) e^{i{\bf n}\cdot {\bf x}}~ c_{\bf n}
    +h_n^{\ast}(\eta) e^{-i{\bf n}\cdot {\bf x}}~ c_{\bf n}^*\,  
\eea 
where a complex number $c_{\bf n}$ can be conveniently presented in its 
polar form:
\be
c_{\bf n} \equiv \rho_{c_{\bf n}} e^{i\phi_{c_{\bf n}}} .
\ee
The gravitational wave solutions at late times are given by 
Eq. (\ref{mumgenLate}). Thus, we have
\bea\label{hcnstat}
h_{\bf n} (\e,{\bf x}) 
&\approx&-{\rho_{c_{\bf n}}\over a(\eta)} \sqrt{2\over \pi}~\Big[ 
\left(A_n \cos y -iB_n \sin y \right) 
e^{i({\bf n}\cdot {\bf x}+\phi_{c_{\bf n}})} \no \\
&&\hspace{0.7in}+ \left(A_n^* \cos y +iB_n^* \sin y 
\right) e^{-i({\bf n}\cdot {\bf x} + \phi_{c_{\bf n}})} \Big]\,.
\eea
If the coefficients $A_n, B_n$ are arbitrary, Eq. (\ref{hcnstat}) is neither
a traveling wave nor a standing wave. A traveling wave is characterized
by two real numbers, namely, an amplitude $A$ and a phase $\phi$; its general 
form is $A \sin (\pm n\eta + {\bf n}\cdot{\bf x} + \phi)$. The minus/plus sign
describes a wave traveling in the positive/negative direction 
defined by the fixed vector ${\bf n}$. A standing 
wave is characterized by three real numbers, {\it viz.}, an amplitude $A$ and 
two phases, and its general form is 
$A \sin(n\eta + \phi_1) \sin ({\bf n} \cdot {\bf x} + \phi_2)$. Different
choices of these free parameters are responsible for concrete space-time 
patterns, but they do not change the wave classification.
 
A little investigation shows that Eq. (\ref{hcnstat}) describes a traveling 
wave if and only if 
\be\label{travelCond}
A_n = \pm B_n \,.
\ee
This constraint can also be written as
\be\label{travelCond1}
A_n = \rho_n e^{i\phi_n} \quad {\rm and} \quad B_n = \pm\rho_n 
e^{i\phi_n} \ \ , 
\ee
where $\rho_n$ and $\phi_n$ are two arbitrary real numbers.
We will refer to this constraint as the {\em traveling wave} condition.
On the other hand,  
Eq. (\ref{hcnstat}) describes a standing wave if and only if 
\be\label{standCond}
A_n = \rho_{A_n} e^{i\phi_n} \quad {\rm and} \quad B_n = \pm i\rho_{B_n} 
e^{i\phi_n} \ \ ,
\ee
where $\rho_{A_n}$, $\rho_{B_n}$, $\phi_n$ are three arbitrary real numbers.
We will refer to this constraint as the {\em standing wave} condition. 
Two special cases of the standing wave condition are when either $A_n=0$ 
or $B_n=0$. This classification of traveling and standing waves was based
on the regime when a given mode satisfies the requirement $y \gg 1$ 
(short-wave regime), but
it can now be applied at earlier times too, when this requirement is not
satisfied (long-wave regime).

\subsection{Stationary versus non-stationary variance}

The variance of a {\it quantized} field (\ref{hij}) is defined by
Eq. (\ref{hmvar}). The essential part of this expression is the power 
spectrum given by Eq. (\ref{power}). At the matter-dominated stage,
the general solution for the gravitational wave mode functions is
represented by Eq. (\ref{mumgen}). One can now find the power spectrum.
We use spherical Bessel functions and replace
the summation over $s$ by the multiplication factor 2. Then,
the general expression for the power spectrum is 
\bea\label{hmvarComp}
h^2(n, \eta) = \frac {2{\cal C}^2 n^2 y^2}{\pi^3 a^2(\eta)}
\left[ |A_n|^2 j_{1}^2(y) +|B_n|^2 j_{-2}^2(y) 
 +2~{\rm Im} \left(A_n^* B_n\right) j_{1}(y) j_{-2}(y)\right] \ \ . 
\eea
 
Using the constraints (\ref{travelCond1}) and (\ref{standCond}), one can 
now specialize the power spectrum to traveling and standing wave cases.  
In the traveling wave case one obtains
\be\label{powerSpectTravel}
h^2(n, \eta) = \frac {2{\cal C}^2 n^2 y^2 \rho_n^2}{\pi^3 a^2(\eta)}
\left[ j_{1}^2(y) + j_{-2}^2(y)\right] \ \ ,
\ee
and in the standing wave case, 
\be\label{powerSpectStand}
h^2(n, \eta) = \frac {2{\cal C}^2 n^2 y^2}{\pi^3 a^2(\eta)}
\left[ \rho_{A_n} j_{1}(y) \pm \rho_{B_n}j_{-2}(y)\right]^2 . 
\ee
When considering the waves that are shorter than the Hubble radius,
$y \gg 1$, one can use asymptotic formulas (\ref{asym}) for the Bessel 
functions. One can also replace $a(\eta)$ with $2 l_H$, since the scale
factor does not practically change during any reasonable observation time. 
Then, in the traveling wave case, the oscillations of the
power spectrum fully disappear, as the oscillating terms $\cos^2 y$ and
$\sin^2 y$ combine to 1. However, the oscillations are most
pronounced in the standing wave case, as Eq. (\ref{powerSpectStand}) 
exhibits a periodic structure
\begin{equation} \label{stand}
h^2(n, \eta) =
\frac {{\cal C}^2 n^2}{2 \pi^3 l_H^2}
\left[ \rho_{A_n} \cos y \pm \rho_{B_n} \sin y\right]^2,~~~y \gg 1 . 
\end{equation}
In other words, the power spectrum of traveling waves is {\em stationary},
whereas the power spectrum of standing waves is {\em non-stationary}.
These two classes of power spectra are subcases of the general situation 
in which 
$A_n = \rho_{A_n} e^{i \phi_{A_n}},~~ B_n = \rho_{B_n} e^{i \phi_{B_n}}$,
and correspondingly,
\begin{equation}
\label{gen}
h^2(n, \eta) =
\frac {{\cal C}^2 n^2}{2 \pi^3 l_H^2}
\left[ \rho_{A_n}^2 \cos^2 y + \rho_{B_n}^2 \sin^2 y +
2 \rho_{A_n} \rho_{B_n} \sin(\phi_{B_n} - \phi_{A_n}) \sin y \cos y \right],~~~
y \gg 1 . 
\end{equation}

At a fixed moment of time, for instance today, the power spectrum
(\ref{stand}) reduces to
\begin{equation} \label{standetaR}
h^2(n, \eta_R) =
\frac {{\cal C}^2 n^2}{2 \pi^3 l_H^2}
\left[ \rho_{A_n} \cos n \pm \rho_{B_n} \sin n\right]^2,~~~n \gg 1 . 
\end{equation}
As one can see, the power spectrum of standing waves contains, quite 
generically, many maxima and zeros at certain $n$'s due to the
oscillatory factors $\cos n,~ \sin n$. The coefficients
$\rho_{A_n}, ~\rho_{B_n}$ are still arbitrary and could, in principle, be 
smooth functions of $n$. However, as we will show in the next section, the
preceding evolution of standing waves through the transition from 
the radiation-dominated era to the matter-dominated era, gives rise to
additional oscillations, namely to oscillations in the 
coefficients $\rho_{A_n}, ~\rho_{B_n}$ themselves, as functions of $n$.

\section{The physical model for relic gravitational wave background}
\label{sec:model}

\subsection{The behavior of the scale factor: Pump field}
\label{subsec:pump}

The matter-dominated ($m$) era with the scale factor (\ref{scale-matter})
was preceded by the radiation-dominated ($e$) era with the scale factor
$a(\eta) \propto \eta$. To simplify the analysis, and without any
essential loss of generality, we assume that the transition from $e$ era 
to $m$ era was instantaneous and took place at some $\eta = \eta_2$. The
redshift of the transition is 
$z_{eq}$: $a(\eta_R)/a(\eta_2) = 1 + z_{eq}$. It is 
believed that $z_{eq}$ is somewhere near $6 \times 10^3$. In its turn, the 
radiation-dominated era was preceded by the initial ($i$) era of expansion, 
whose nature and
scale factor are, strictly speaking, unknown. To simplify the analysis,
and since the wave equations admit simple exact solutions in case of 
power-law scale factors, we assume that the $i$ era, similar to the
$e$ and $m$ eras, was also described by a power-law scale factor.
We parametrize the $i$ era by $a(\eta) \propto |\eta|^{1+\beta}$
(compare with \cite{LPG1}). The
transition from $i$ era to $e$ era takes place at some $\eta = \eta_1$
and at redshift $z_i$: $a(\eta_R)/a(\eta_1) = 1+ z_{i}$. Further analysis
shows (see below) that in order to get the right amplitude of the generated
perturbations, the numerical value of $z_i$ should be
somewhere near $10^{30}$. 

We now write the full evolution of the growing scale factor explicitly: 
\be
   a(\eta )= l_o | \eta |^{1+\beta}, \quad
   \eta \leq \eta_1,~~\eta_1 <0,~~\beta < -1 \ \ ,
\ee
\be
    a(\eta )= l_o a_e (\eta - \eta_e) , \quad
    \eta_1 \leq \eta \leq \eta_2 \ \ ,
\ee
\be
   a(\eta )=  2l_H (\eta - \eta_m )^2 , \quad \eta_2 \leq \eta . 
\ee
The continuous joining of $a(\eta)$ and $a^{\prime} (\eta)$ at the 
transition points fully determines all the participating constants
in terms of $l_H,~z_i,~z_{eq}$ and $\beta$. Concretely,
\bea \label{etaconst}
\eta_R &=& 1 + \eta_m, ~~ 
\eta_m = -\frac{1}{2\sqrt{1+z_{eq}}}
	\left[1-\beta \frac{1+z_{eq}}{1+z_i}\right],~~\ \
\eta_2= \frac{1}{2\sqrt{1+z_{eq}}}
	\left[1+\beta \frac{1+z_{eq}}{1+z_i}\right], \no\\ 
\eta_e &=& \frac{1}{2} \beta \frac{\sqrt{1+z_{eq}}}{1+z_i},~~
\eta_1= \frac{1}{2}(1+ \beta) \frac{\sqrt{1+z_{eq}}}{1+z_i},
\eea
and
\begin{equation} \label{aconst}
l_o a_e= \frac{4 l_H}{\sqrt{1+z_{eq}}},~~
l_o=l_H \frac{2^{2+\beta}}{|1+\beta|^{1+\beta}}\frac{(1+z_{eq})^
{-(\beta+1)/2}}{(1+z_i)^{-\beta}}.
\end{equation}
The case $\beta = -2$ is known as the de Sitter inflation. In this particular
case,
\begin{equation} \label{loinfl}
l_o|_{\beta= -2} =l_H \frac{\sqrt{1+z_{eq}}} {(1+z_i)^2}.
\end{equation}

For the CMB calculations we will also need the redshift $z_{dec}$ of the 
last scattering surface $\eta = \eta_E$ (with the subscript $E$ denoting 
``emission''), where the CMB photons have 
decoupled from rest of the matter: $a(\eta_R)/a(\eta_E) = 1+ z_{dec}$. 
The numerical  value of $z_{dec}$ is somewhere near 1000. The time of
decoupling $\eta_E$ is 
\be \label{etaE}
\eta_E = \frac{1}{\sqrt{1+z_{dec}}}- \frac{1}{2\sqrt{1+z_{eq}}} + 
 \beta \frac{\sqrt{1+z_{eq}}}{2(1+z_i)}. 
\end{equation}

All the formulas above are exact and we will be using them often, but 
surely one can get an excellent approximation by neglecting 1
in comparison with $z_{i},~z_{eq}$ and $z_{dec}$.

\subsection{Squeezed gravitational waves}
\label{subsec:sqGW}

As soon as the scale factor and initial conditions for the mode functions
are strictly defined, the coefficients $A_n, B_n$ 
in Eq. (\ref{mumgen}) are strictly calculable \cite{LPG93,AK}. 
The general solution for $\mu_n$ at the $e$ stage is
\be \label{mue}
\mu_n(\eta) = B_1 e^{-in(\eta - \eta_e)} +  B_2 e^{in(\eta - \eta_e)},
\ee
but the preceding evolution allows one to specify $B_1$ and $B_2$. The
waves subject to amplification at the $i$ stage have the wave numbers $n$
satisfying the condition $n|\eta_1| \ll 2 \pi |1+\beta|$, that is,
\[
\frac{n}{n_H} \ll \frac{1+z_i}{\sqrt{1+ z_{eq}}}.
\]
For these waves, one finds 
\be \label{B}
B_1 \approx - B_2 \approx F(\beta) \left(\frac{n \sqrt{1+z_{eq}}}{1+z_i}
\right)^{\beta} \equiv B~,
\ee
where  
\be \label{F}
F(\beta) = - e^{i(x_0 + \pi \beta/2)}\frac{ \sqrt{\pi}|1+\beta|^{\beta+1}}
{2^{2(\beta+1)} \Gamma(\beta+3/2) \cos~ \beta\pi} \ \ ,
\ee
and $x_0 \equiv n\eta_0$ \cite{LPG93}.
Note that in the particular case $\beta = -2$, 
\be \label{|B|2}
|B|^2 = \frac {4(1+z_i)^4}{n^4 (1+z_{eq})^2},~~~~~\beta =-2. 
\ee

The fact that $B_1 \approx -B_2$ demonstrates that the gravitational wave
modes ${\bf n}$ are (almost) standing waves at the $e$ stage. These 
standing waves are encountering
the nonzero barrier $a^{\prime \prime}/a$ at the $m$ stage. To find the
coefficients $A_n, B_n$ in Eq. (\ref{mumgen}) one needs to join continuously
the $\mu_n(\eta)$ and $\mu_n^{\prime} (\eta)$ at the transition point
$\eta = \eta_2$. This calculation allows one to find the coefficients 
$A_n, B_n$, but it also shows (as expected) that the standing-wave
character of the field at the $e$ stage leads to the appearance of 
oscillations in the coefficients $A_n,~B_n$ of the field at the
$m$ stage. This is a general phenomenon which we will also discuss 
in connection with density perturbations. Explicitly,
\begin{mathletters}
\label{C1C2}
\bea 
A_n & =&  -i \sqrt{\frac{\pi}{2}} \frac {B}{4 y_2^2}\left[(8y_2^2 -1)\sin y_2 +
4y_2 \cos y_2 +\sin 3y_2 \right], \\
B_n & = & - \sqrt{\frac{\pi}{2}} \frac {B}{4 y_2^2}\left[(8y_2^2 -1)\cos y_2- 
4y_2 \sin y_2 +\cos 3y_2 \right], 
\eea 
\end{mathletters}
where
\[
y_2 = n(\eta_2 -\eta_e) = \frac{n}{n_m},~~{\rm and}~~n_m= 2\sqrt{1+z_{eq}}.
\]
The numerical value of $n_m$ is about 160. This corresponds to wavelengths
that are 15 times shorter than $l_H$. Clearly, the coefficients
(\ref{C1C2}) satisfy the standing wave condition (\ref{standCond}). It is 
easy to check that had one artificially chosen traveling waves at the $e$ 
stage, by assuming that either $B_1=0$ or $B_2 = 0$, the oscillations in
$A_n,~B_n$ would have been suppressed.
     
The asymptotic expressions for $A_n,~B_n$ are as follows.  
For relatively short waves, i.e., $y_2 \gg 1,~n\gg n_m$, one has
\bea
\label{shortC}
A_n \approx -i2 \sqrt{\pi/2}B \sin y_2,~~ B_n 
\approx -2 \sqrt{\pi/2}B \cos y_2.
\eea
Using these coefficients in Eq. (\ref{mumgenLate}), one finds the 
$\mu_n(\eta)$ at the $m$ stage, for $n \gg n_m$:
\[
\mu_n(\eta) \approx - i 2B \sin[n(\eta -\eta_e)]. 
\] 
Of course, this function is simply a continuation of the relevant  
standing wave solution 
($B_1 \approx -B_2 \approx B$) at the $e$ stage,  Eq. (\ref{mue}), to
the $m$ stage.
Indeed, the height of the barrier at the $m$ stage is
\[
\frac{a^{\prime \prime}}{a}\Big|_{\eta= \eta_2} = \frac{1}{2} n_m^2,
\]
so the waves with $n \gg n_m$ stay above the barrier and experience no
changes. However, the waves with $n \approx n_m$ and $n \ll n_m$ are 
affected by the barrier.  
For relatively long waves, i.e., $n\ll n_m$, one has 
\bea\label{longC}
A_n \approx -i \frac{3 \sqrt{\pi}}{2 \sqrt 2} B y_2^{-1}, ~~ 
B_n \approx \frac{8 \sqrt{\pi}}{45 \sqrt{2}} B y_2^4 \ \ ,  
\eea
such that $|B_n| \ll |A_n|$. The formulas above are in full agreement 
with \cite{LPG93} if one takes into account the change of notations:
$A_n = \sqrt{\pi/2} C_{1n},~B_n = i \sqrt{\pi/2} C_{2n}$. In particular,
the long-wavelength part of the power spectrum at the time of decoupling
$\eta = \eta_E$ is given by
\be \label{powgw}
h^2 (n, \eta_E) \approx {4 \over \pi} {l_{Pl}^2 \over l_H^2} (1+ z_{eq})
n^4 |B|^2, ~~~ n \ll n_{dec}=  \sqrt{1+z_{dec}} \,.
\ee
This part of the spectrum is primordial, in the sense that it has not
changed since the beginning of amplification. 
In the particular case $\beta = -2$, and using Eq. (\ref{|B|2}),
we obtain the flat (independent of $n$) primordial spectrum  
\be \label{powgw2}
h^2 (n, \eta_E) \approx {16 \over \pi} {l_{Pl}^2 \over l_H^2} {(1+ z_i)^4 
\over (1+z_{eq})}, ~~~ \beta = -2, ~~~ n \ll n_{dec}= \sqrt{1+z_{dec}} \,.
\ee

\begin{figure}[!hbt] 
\caption{Plot of $g^2(x,b=5)$ versus $x$.} 
\centerline{}
\centerline{}
\centerline{}
\centerline{}
\centerline{\psfig{file=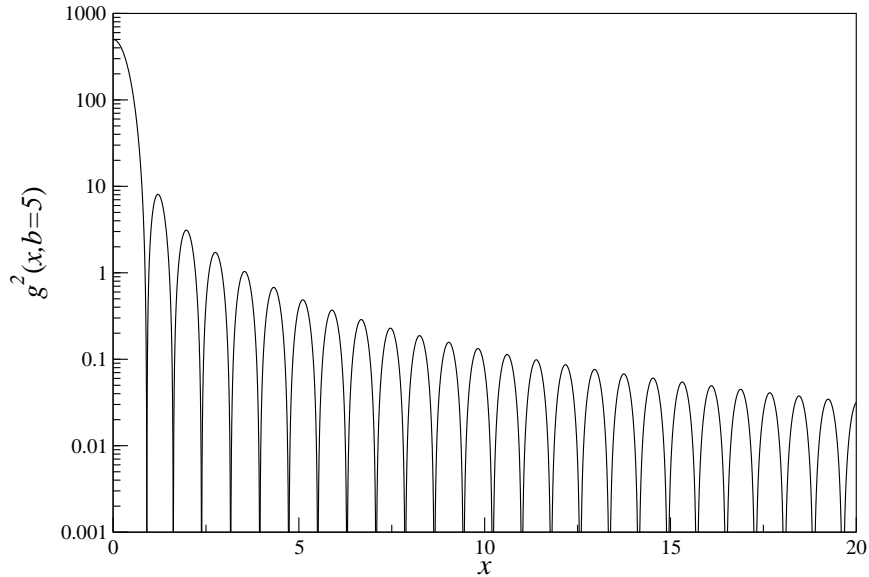,height=3.in,width=4.5in}}
\label{fig:gSquare}
\end{figure}

In preparation for the discussion of CMBR anisotropies, we show in
Fig. \ref{fig:gSquare} the numerically calculated spectrum $h^2(n, \eta_E)$, 
including the beginning of its oscillations. We use the notations
\[
x \equiv y_2 = {n \over n_m}, ~~~~ b \equiv {n_m \over n_{dec}}=
{2 \sqrt{1+z_{eq}} \over \sqrt{1+z_{dec}} }\,.
\]
The substitution of $A_n, B_n$ given by Eq. (\ref{C1C2}) into the general
expression (\ref{hmvarComp}) produces the exact power spectrum
\be \label{powexact}
h^2 (n, \eta_E) = {l_{Pl}^2 \over 4 \pi l_H^2}(1+z_{dec}) n^4 |B|^2 g^2 (x, b)
\ \ ,
\ee
where
\[
g^2(x,b) = \left[\rho_{1}(x) j_{1}(bx) - \rho_{2}(x) j_{-2} (bx)\right]^2,~~~
x \equiv {n \over n_m}, ~~~ b \equiv {n_m \over n_{dec}},
\]
\[
\rho_{1}(x) = \frac {1}{x^2}\left[(8x^2 -1)\sin x + 4x \cos x +\sin 3x \right],
\]
\[ 
\rho_{2}(x) = \frac {1}{x^2}\left[(8x^2 -1)\cos x- 4x \sin x +\cos 3x \right]. 
\]
Note that the rigorously evolved mode functions single out the lower sign
in the general formula (\ref{powerSpectStand}) for standing waves. 
Fig. \ref{fig:gSquare} shows the function $g^2(x,b)$ for $b = 5$. 

The positions of maxima and zeros of the power spectrum are well 
approximated by the short-wavelength limit of $g^2(x,b)$:
\[
g^2(x,b) \approx {64 \over b^2 x^2} \sin^2[(b-1)x].
\]
In the case of $b = 5$, the maxima and zeros of the function $\sin^2 4x$ 
are ordered as follows:
\be \label{gmax}
{\rm Maxima:}~~ x_k^{max} = {\pi \over 8}(2k+1),~~~x_k^{max} = x_0^{max}(2k+1),
~~~k = 0,1,2,3,...,
\ee
\be \label{gmin}
{\rm Zeros:}~~ x_k^{min} ={\pi \over 8} 2k,~~~x_k^{min}= x_1^{min}k,
~~~k = 1,2,3,....
\ee
The analytical forecast (\ref{gmax}), (\ref{gmin}) shifts the first 
few features to the left (smaller
$x$) as compared with the more accurate numerical calculation, but the forecast 
becomes progressively more accurate for later features. For example, the 
analytical formula predicts $x_1^{min} = 0.79,~ x_1^{max} = 1.18,~ 
x_8^{min} = 6.28,~ x_8^{max} = 6.68$, whereas the numerical calculation
places these features at $x_1^{min}= 0.91,~ x_1^{max} =1.21,~ 
x_8^{min} = 6.29,~ x_8^{max} = 6.68$. The zero-th maximum $x_0^{max}$
would be placed, according to the analytical formula (\ref{gmax}), at
$x_0^{max}= 0.393$. The crucial rule, however, is that the positions of 
minima, starting from $x_1^{min}$, are ordered in the 
proportion $1:2:3:4..$,
whereas the positions of maxima, starting from the zero-th
``would-be" maximum, are ordered in proportion $1:3:5:7...$. 
We will see in Sec. \ref{sec:anisotropies} how these 
features are being reflected in the oscillations of the $C_l$ multipoles.

\section{Distinguishing between the presence and the absence of squeezing}
\label{sec:squeeze}

As demonstrated above, the quantum-mechanically generated (squeezed) 
gravitational waves
form a non-stationary background, whose power spectrum is fully determined
by fundamental constants and parameters of the gravitational pump field.
In this section we show that the hypothesis whereby the gravitational
wave background is postulated to be stationary, and therefore 
non-squeezed, is in conflict with some other cosmological data.
Since we concentrate on stationarity versus non-stationarity, the comparison
of the two backgrounds should be fair, in the sense that their today's
(for $\eta = \eta_R$) band-powers should be assumed equal. However, 
as we will show below, when one returns 
back in time to, say, the decoupling era, the alternative background is bound 
to have too much power in long-wavelength perturbations. The further
extrapolation back in time destroys the usual (and partially tested) 
assumption that the cosmological perturbations remain small all the way 
down to the nucleosynthesis era and beyond. In Sec. \ref{sec:anisotropies},  
we will also show that the alternative background does not produce 
oscillations in the $C_l$ multipoles, whereas the physical background does.    

The power spectrum of the physical background is given by the general
expression (\ref{hmvarComp}) where $A_n, B_n$ are determined 
by Eq. (\ref{C1C2}). The substitution of Eq. (\ref{C1C2}) in 
Eq. (\ref{hmvarComp}) results in formula (\ref{powerSpectStand})
(with the lower sign) where $\rho_{A_n}, \rho_{B_n}$ are absolute values
of the coefficients (\ref{C1C2}). On the other hand,  the general form of 
the power spectrum for the alternative (stationary) background is given 
by Eq. (\ref{powerSpectTravel}), where $\rho_n$ should be found from the
fair comparison:  
\be\label{fairComp2}
\rho_n^2\left[ j_{1}^2(n) + j_{-2}^2(n)\right]=
\left[\rho_{A_n} j_{1}(n) - \rho_{B_n} j_{-2}(n)\right]^2. 
\ee
We will find $\rho_n$ using asymptotic expressions (\ref{shortC}),
(\ref{longC}) in different intervals of $n$. 
In the region $n \gg n_m$, formula (\ref{shortC}) yields  
\be \label{shortrho}
\rho_{A_n} \approx \sqrt{2 \pi} |B| \left|\sin \frac{n}{n_m}\right|, ~~ 
\rho_{B_n} \approx \sqrt{2 \pi} |B|\left|\cos \frac{n}{n_m}\right|.  
\ee
In the region $n \ll n_m$ formula (\ref{longC}) yields to
\be \label{longrho}
\rho_{A_n} \approx \frac{3 \sqrt{\pi}}{2 \sqrt 2} |B| \frac{n_m}{n}, ~~ 
\rho_{B_n} \approx \frac{8 \sqrt{\pi}}{45 \sqrt{2}} |B| 
\left(\frac{n}{n_m}\right)^4.
\ee  
 
Let us start from $n \gg n_m$. Using asymptotic expressions (\ref{asym})
for the spherical Bessel functions, and replacing $\sin^2 n$ with $1/2$ in
the right-hand-side of (\ref{fairComp2}), one finds
\be \label{rho1}
\rho_n^2 \approx \pi |B|^2, ~~~n \gg n_m.   
\ee
Now turn to the interval $n_m \gg n \gg 1$. The $\rho_{A_n},~\rho_{B_n}$ are
now given by (\ref{longrho}). One can still use the asymptotic
formulas (\ref{asym}), but the second term in the right-hand-side of
(\ref{fairComp2}) is much smaller than the first one and therefore can be
neglected. Then, one derives
\be \label{rho2}  
\rho_n^2 \approx \frac{9 \pi}{16} |B|^2 \frac{n_m^2}{n^2},~~ n_m \gg n \gg 1.
\ee
Finally, in the region $n \ll 1$, one uses asymptotic formulas
$j_1(n) \sim n/3, ~~j_{-2}(n) \sim 1/n^2$. In either side of 
(\ref{fairComp2}), the first term is smaller than the
second term, and can be neglected. Then, one obtains
\be \label{rho3}  
\rho_n^2 \approx \frac{\pi}{8} |B|^2 n_m^2 n^4,~~  n \ll 1.
\ee
These formulas give a piece-wise representation for the smooth alternative
spectrum which today has, in all intervals of $n$, approximately the same 
power as the physical spectrum does.  

Since the coefficients $\rho_{A_n},~\rho_{B_n},~\rho_n$ are fully determined,
one can now derive the forms of the two power spectra at other times. 
We will compare the two backgrounds at the time of decoupling
$\eta = \eta_E$. One needs to consider formulas (\ref{powerSpectTravel}) and  
(\ref{powerSpectStand}), where $a(\eta) = a(\eta_E)$ and $y= y_E$,
\be \label{yE}
y_E = n(\eta_E - \eta_m)= \frac{n}{n_{dec}}= \frac{n}{\sqrt{1+z_{dec}}}.   
\ee
Clearly, in the band of sufficiently short waves, $n \gg \sqrt{1+z_{dec}}$, 
both spectra increase power in the same proportion, simply as a result of 
changing of $a(\eta)$ from $a(\eta_R) = 2 l_H$ to the smaller value 
$a(\eta_E) = 2 l_H/(1+z_{dec})$. So, in this range of $n$ the ratio 
of powers in the two backgrounds is 1:   
\be \label{ratio1}
\frac{h^2(n, \eta_E)|_{alt}}{h^2(n, \eta_E)|_{phys}} = 1,~~n \gg \sqrt{1+z_{dec}}.
\ee
However, this ratio is significantly larger than 1 in longer waves. This is
seen from the general formula
\be \label{ratiogen}
\frac{h^2(n, \eta_E)|_{alt}}{h^2(n, \eta_E)|_{phys}} = 
\frac {\rho_n^2\left[ j_{1}^2(n/n_{dec}) + j_{-2}^2(n/n_{dec})\right]} 
{\left[\rho_{A_n} j_{1}(n/n_{dec}) - \rho_{B_n} j_{-2}(n/n_{dec})\right]^2} 
\ee
applied to longer waves. Indeed, in the interval $1 \ll n \ll \sqrt{1+z_{dec}}$
(and, hence, $n \ll n_m$) one uses the small argument approximation for the
Bessel functions, neglects second terms in numerator and denominator, 
and takes $\rho_n^2$ from Eq. (\ref{rho2}). This calculation results in
\be \label{ratio2}
\frac{h^2(n, \eta_E)|_{alt}}{h^2(n, \eta_E)|_{phys}} = 
\frac{9}{2} \frac{(1+z_{dec})^3}{n^6},~~1 \ll n \ll \sqrt{1+z_{dec}}.
\ee
This ratio is comparable with 1 only at 
$n \sim \sqrt{1+z_{dec}}$ where (\ref{ratio2}) goes over into (\ref{ratio1}). 
But the ratio (\ref{ratio2}) is much larger than 1 for smaller $n$'s. 
As for the region $n \ll 1$, one applies the same approximations as in the 
previous case, but takes  
$\rho_n^2$ from Eq. (\ref{rho3}). This gives the result
\be \label{ratio3}
\frac {h^2(n, \eta_E)|_{alt}}{h^2(n, \eta_E)|_{phys}} = 
(1+z_{dec})^3,~~n \ll 1.
\ee
This ratio is universally (independently of $n$ in this region) much 
larger than 1. 

It is easy to understand these results. The parts of the general solution
(\ref{mumgen}) with $j_1(y)$ and $j_{-2}(y)$ are called, respectively,
the growing and decaying solutions. This classification reflects their
different behavior in the small argument approximation for the Bessel
functions. The growing and
decaying solutions are necessarily present in the power spectra of both
backgrounds. However, 
the stationary background contains $j_1^2(y)$ and $j_{-2}^2(y)$ always 
with equal coefficients, whereas the non-stationary background contains 
$j_1(y)$ and $j_{-2}(y)$ with equal (up to oscillations) coefficients 
(\ref{shortrho}) only
for sufficiently short waves. In longer waves, the decaying solution 
of the stationary background becomes progressively more important when one
goes back in time. Since the physical background is expected to
have enough power to produce the lower order anisotropy in the CMB temperature
at the level $\delta T/ T \sim 10^{-5}$, the alternative
background would have produced this anisotropy at the
unacceptably high level $\delta T/ T \sim 3 \times 10^{-1}$.

\section{Microwave background anisotropies caused by relic gravitational waves}
\label{sec:anisotropies}

The key element in formula for the temperature variation $\delta T/ T$
seen in a given direction ${\bf e}$ \cite{sw} is the $\eta$-time derivative
of metric perturbations evaluated along the CMB photon's path between the event
of reception (R) and the event of emission (E):
\be \label{sw}
\frac {\delta T}{T}({\bf e}) = \frac {1}{2} \int_0^{w_1} 
\left[\frac{\partial h_{ij}}{\partial \eta} e^i e^j \right]_{path} dw.
\ee
The upper limit of integration is
\[
w_1 = \eta_R - \eta_E = 1 - \frac {1}{\sqrt{1+ z_{dec}}},
\]
and the integration is performed along the path $\eta = \eta_R - w,~
{\bf x} = {\bf e} w$. In the case of density perturbations, the integral in 
formula (\ref{sw}) should be augmented by the additive term representing
initial conditions: an intrinsic variation of temperature 
at $\eta= \eta_E$ and a possible velocity of the last scattering 
electrons with respect to the chosen coordinate
system, which is synchronous and comoving with the perturbed, gravitationally 
dominant pressureless matter, possibly -- cold dark matter (CDM).    

Similar to the perturbation field $h_{ij}$ itself, the temperature variation
$\delta T /T$ is also a quantum-mechanical
operator. To establish contact with macroscopic physics, we need to calculate
the correlation function
\[
\left\langle 0\left| {\delta T \over T}
   ({\bf e}_1) {\delta T \over T} ({\bf e}_2)\right|0 \right\rangle. 
\] 
We use the mode functions (\ref{mumgen}) and the normalization constant
${\cal C} = \sqrt{16 \pi} l_{Pl}$. Then, it can be shown \cite{LPG93} that
the correlation function takes the elegant form 
\begin{equation}\label{Tmom}
   \left\langle 0\left| {\delta T \over T}
   ({\bf e}_1) {\delta T \over T} ({\bf e}_2)\right|0 \right\rangle
 = l_{Pl}^2 \sum_{l=2}^\infty K_l P_l(\cos\delta ),
\end{equation}
where $P_l(\cos\delta )$ are Legendre polynomials for the separation angle 
$\delta$ between the unit vectors ${\bf e}_1$ and  ${\bf e}_2$, and 
\[
K_l = (2l+1)(l-1)l(l+1)(l+2)~F_l, 
\]
where 
\begin{equation}\label{Fl}
   F_l = \int_0^\infty n^2 \left| \int_0^{w_1}
   {J_{l+1/2} (nw) \over (nw)^{5/2}}
   f_n(\eta_R-w)dw \right|^2 dn, 
\end{equation}
and
\be
f_n(\eta_R-w) = {1\over\sqrt{2n}} 
{\left(\frac{\mu_n}{a}\right)^\prime}\Big|_{\eta = \eta_R - w}. 
\ee
We will also be using the multipole moments $C_l$ defined by
\[
l_{Pl}^2 K_l = \frac{2l+1}{4 \pi} C_l.
\]

The central quantity for the calculation of $F_l$ is the function
\begin{equation} \label{muprime} 
   \left( {\mu_n \over a} \right)^\prime
 = -{n\over a} \sqrt{y}
    (A_n J_{5/2}(y) +i B_nJ_{-5/2}(y)).
\end{equation}
To get more insight into $F_l$ we introduce the two new functions: 
$\psi_{\pm l}$, defined by the respective integrals:
\be
   \psi_{\pm l} (nw_1) = \int_0^{nw_1}dx \frac {J_{l+1/2} (x)}{x^{5/2}}
\frac{J_{\pm 5/2}(n-x)}{(n-x)^{3/2}}  .
\ee
Then, the general expression for $F_l$ takes the form
\be\label{Flgen}
F_l =  {1\over 8 l_H^2 } \int_0^\infty dn~n^5 
\left[ |A_n|^2 {\psi}_l^2 +|B_n|^2 {\psi}_{-l}^2 +2{\rm Im} 
(A_n^* B_n) {\psi}_l {\psi}_{-l} \right]  . 
\ee
This expression simplifies when the coefficients $A_n, B_n$ represent
traveling (\ref{travelCond}) or standing (\ref{standCond}) waves.
For a stationary background one gets
\be\label{Fltrav}
F_l =  {1\over 8 l_H^2 } \int_0^\infty dn~n^5 
\rho_n^2\left[ {\psi}_l^2 + {\psi}_{-l}^2 \right]  ,
\ee
while for the physical nonstationary background one gets  
\be\label{Flstand}
F_l =  {1\over 8 l_H^2 } \int_0^\infty dn~n^5 
\left[ \rho_{A_n} {\psi}_l - \rho_{B_n} {\psi}_{-l} \right]^2  . 
\ee
These formulas explain the different behavior of the multipole moments $C_l$
in the two cases. We demonstrate this with the help of numerical calculations.

\begin{figure}[!hbt] 
\caption{The solid line depicts the plot of $l(l+1)C_l$ versus 
$l$ in the physical model, normalized 
such that at $l=10$, we have $l(l+1)C_l = 6.4\times 10^{-10}$, which  
tallies with observations. The (red) dashed line is the corresponding plot 
in the alternative model.  
Here, we take $\beta = -2$, and the redshifts at 
$\eta_2$ and $\eta_E$ to be $z_{eq} = 10000$ and
$z_{dec} = 1000$, respectively.}
\centerline{}
\centerline{}
\centerline{}
\centerline{}
\centerline{\psfig{file=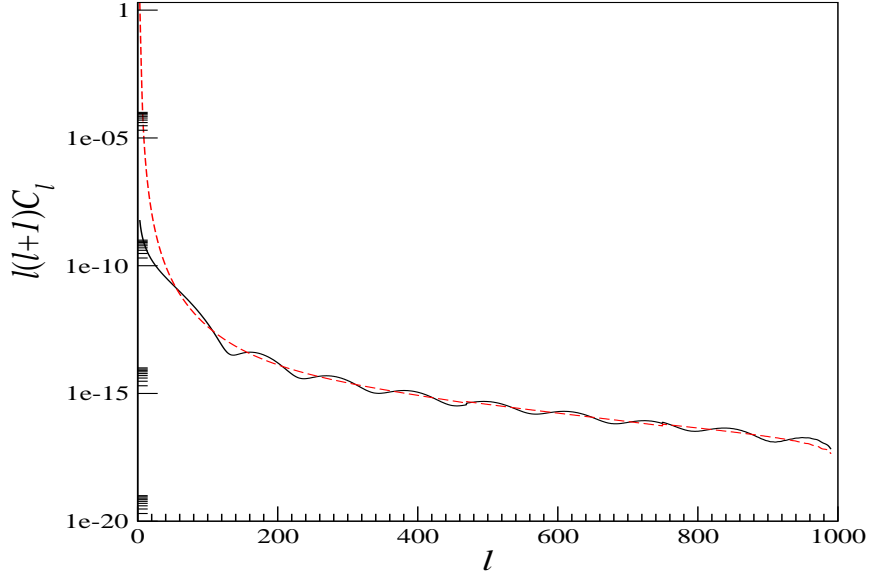,height=3.in,width=4.5in}}
\label{fig:Cl}
\end{figure}

In Fig. \ref{fig:Cl} we show by a solid line the graph of the function 
$l(l+1) C_l$ calculated with the help of Eq. (\ref{Flstand}) and
Eq. (\ref{C1C2}). The cosmological parameters were chosen, for illustration, 
as $z_{eq} = 10^4,~ z_{dec}= 10^3,~ \beta = -2$. The parameter $z_i$ is 
adjusted in such a manner ($z_i = 10^{29.5}$) that the graph goes through the 
point $l(l+1)C_l =
6.4 \times 10^{-10}$ at $l=10$, which agrees with observations. 
Our attention is focused, however, on the oscillations in this function.
The dashed line shows the same function for the alternative
stationary background. The cosmological parameters are the same as above, 
but the calculation is performed with the help of Eq. (\ref{Fltrav}) and
the coefficients $\rho_n$, found from the 
condition (\ref{fairComp2}) of 
the fair comparison. The remarkable (even if expected) result is that 
the stationary background of gravitational waves does not produce 
oscillations in the angular power spectrum $C_l$, whereas the non-stationary
background does.

The numerical positions of minima and maxima in the oscillating graph are
ordered as follows:
\[
{\it Minima}:~ l_1 = 137,~ l_2 =237,~ l_3= 344,~ l_4=456,~ 
l_5= 569,~ l_6 =682,~ l_7 =796,
\]
\[
{\it Maxima}:~ l_1=161,~ l_2 =269,~ 
l_3=381,~ l_4 =494,~ l_5 =609,~ l_6= 723,~ l_7 =839. 
\]
Clearly, these features
reflect the oscillations in the metric power spectrum. Judging from the
mathematical structure of the participating Bessel functions, it is likely
that the positions of features in the $n$-space are related to the
positions of features in the $l$-space by a simple numerical 
coefficient of order 1. It is difficult to find out this coefficient 
analytically, though. Remembering that the positions of first
features may be displaced by 10-15 percent, as compared with the 
analytical forecast (\ref{gmin}), (\ref{gmax}),
we put the zero-th ``would be" maximum at $l_0^{max} = 56$. Then, our 
simple analytical formula places the next features in the following
positions: $l_1^{min}= 112,~ l_1^{max}= 168,~ l_2^{min} = 224,~ l_2^{max}=280,~ 
l_3^{min}= 336,~ l_3^{max}= 392,~ l_4^{min}= 448,~ l_4^{max}= 504,~ 
l_5^{min}= 560,~ l_5^{max}= 616,~ l_6^{min}= 672,~ l_6^{max}= 728,~ 
l_7^{min}= 784,~ l_7^{max}= 840$. Comparing this prediction with
the numerically calculated positions, we find them in a fairly good
agreement.

This investigation of gravitational waves provides us with a guidance
for the technically more complicated case of density perturbations.  

\section{Density perturbations and the $C_{\ell}$ features}
\label{sec:dp}

The general expression for metric perturbations, associated with density
perturbations, is given by Eq. (\ref{hij}) with the polarization tensors 
explained in the Introduction. When one is actually writing down the perturbed
Einstein equations, it turns out that it is 
more convenient to work with the mode functions $h(\eta)$ and $h_l(\eta)$
instead of the original mode functions ${\stackrel{s}{h}}_n (\eta )$. The
relationship between them is: 
\be \label{scallong}
{\stackrel{1}{h}}_n (\eta )= 
\sqrt{\frac{3}{2}}\left(h(\eta)- \frac{1}{3} h_l(\eta)\right),~~~ 
{\stackrel{2}{h}}_n (\eta )= \frac{1}{\sqrt{3}} h_l(\eta), 
\ee
where the wave-number index $n$ on the mode functions $h(\eta)$ and $h_l(\eta)$
is implicit. The function $h(\eta)$ is the purely scalar part of perturbations,
it enters $h_{ij}$ with the polarization structure
${\stackrel{1}{P}}_{ij} = \delta_{ij}$, whereas the function $h_l(\eta)$ is
the purely longitudinal-longitudinal part of perturbations, it enters $h_{ij}$
with the polarization structure ${\stackrel{2}{P}}_{ij} = -n_i n_j/ n^2$.
We will follow the same strategy as in the case of gravitational waves, and will
start from exact solutions to the perturbed Einstein equations in different 
cosmological eras. 

\subsection{Density perturbations at the matter-dominated stage}

The matter-dominated stage is driven by a pressureless matter; possibly, - 
cold dark matter (CDM).
The general solution to the perturbed equations at the $m$ stage
can be simplified by using the available freedom within the class of synchronous
coordinate systems. By using this freedom, one specializes to the
unique coordinate system, which is synchronous and comoving with the perturbed
pressureless matter. In this coordinate system, the general solution is  
\be \label{genm}
h(\eta) = C_1,~~h_l(\eta)=\frac{1}{10}C_1 n^2 (\eta-\eta_m)^2 -
\frac{1}{3}C_2 \frac{(\eta_2-\eta_m)^3}{(\eta - \eta_m)^3},
\ee
where $C_1, C_2$ are arbitrary complex numbers. The matter 
density perturbation is
\be \label{enden}
\frac{\delta \epsilon}{\epsilon_0} = \frac{1}{2} h_l(\eta),
\ee
and the velocity $v^j$ of matter elements, including perturbations, 
\be \label{vel}    
\frac{v^j}{c} = \frac{T_0^j}{T_0^0}, 
\ee
is by construction zero, i.e.,
\be \label{velz}    
\frac{v^j}{c} = 0. 
\ee
The above solution is well known since the times of E. M. Lifshitz and can be 
found in various publications, up to possible misprints. For instance, this
solution follows from equations given in Ref. \cite{hswz}, if one corrects for 
a misprint in the last line of Eq. (A8): the second term there should 
actually enter with coefficient 2. 

The adopted choice of the unique coordinate system, which is both 
comoving and synchronous,
not only greatly simplifies the form of solutions, but is also
needed for a proper formulation of the $\delta T/T$ calculations. As long as
the emitter and the receiver are riding on the time-like geodesics
$x^i = const.$ of this perturbed metric, the Sachs-Wolfe integral   
(\ref{sw}) is the full answer; there are no extra velocity 
contributions to this integral. The additive velocity contributions arise 
only if the emitter or the receiver are moving with respect to this unique 
coordinate system, that is, when they are not described by the world-lines
$x^i =const.$. 

For the growing solution (namely, the terms with coefficient $C_1$), the 
Sachs-Wolfe integral
can be taken exactly. It appears that astrophysical literature calls by 
gravitational ``Sachs-Wolfe effect" only a part of what is actually 
contained in the Sachs-Wolfe paper \cite{sw}. Invariably, by the 
``Sachs-Wolfe effect" are meant only two terms, which are, 
roughly speaking, the difference of ``gravitational potentials" at the events
of emission (E) and reception (R). Two other terms in their full 
formula (43), which are the difference of the ``gradients of the 
gravitational potential", are being systematically ignored. Possibly, 
this happened because Sachs and Wolfe addressed one of these gradient 
terms in the words: ``this second term is normally
small". This second term is indeed small for small wave-numbers, but it is 
in fact dominant for large wave-numbers, which are responsible for the
dipole $C_1$ and for the $C_l$ multipoles near the peak at $l \sim 200$. 
For example, correct implementation of the full Sachs-Wolfe formula (43) for 
calculation of the dipole $C_1$ gives a number that is 5 orders
of magnitude greater than the number following from the ``Sachs-Wolfe 
effect" counterpart of the full formula. The lack of ergodicity on
a 2-sphere provides a 1$\sigma$ uncertainty in the $C_l$'s, roughly
at the level $\Delta C_l \simeq \sqrt{2/(2l+1)} C_l$. 
We say ``roughly" because the statistic of the underlying random variable 
is not Gaussian, it is described by the product of an exponent and
the modified Bessel function $K_0$ \cite{gristat}. 
In the case of the dipole $C_1$, the uncertainty amounts 
to $\Delta C_1/C_1 \approx 0.8$.
Clearly, this factor-of-two uncertainty cannot cover the 5 orders
of magnitude disparity in the results; quite simply, 
the result based on the misinterpreted ``Sachs-Wolfe effect" is what is 
wrong. For details, see Ref. \cite{dg}.    
 
The coefficients $C_1, C_2$ in the general solution (\ref{genm}) are, 
so far, arbitrary, but they are determined
by the previous evolution of density perturbations ($d.p.$).

\subsection{Density perturbations at the radiation-dominated stage} 

The ``master equation" at the $e$ stage is
\be \label{emast}
\nu^{\prime\prime} + \frac{1}{3} n^2 \nu =0,
\ee
where the coefficient $1/3$ enters because we have used $c_l/c = 1/\sqrt{3}$,
which is valid deep in the radiation-dominated era.
By the time of decoupling, the plasma sound speed decreases slightly
below this value, depending on the baryon content, and we will 
account for this fact by returning 
back to $c_l/c$ in appropriate places. The general solution 
to Eq. (\ref{emast}) is always oscillatory as a function of time:
\be \label{enu}
\nu =B_1 e^{-i\frac{n}{\sqrt{3}}(\eta- \eta_e)} + 
B_2 e^{i\frac{n}{\sqrt{3}}(\eta- \eta_e)},
\ee
where $B_1, B_2$ are arbitrary complex numbers. All the metric and matter
perturbations can now be found from solutions (\ref{enu}). For metric 
perturbations, one has:  
\be \label{eh}
h(\eta) = \frac{a^{\prime}}{a^2} \left[\int_{\eta_1}^{\eta} \nu {\rm d}\eta + 
C_e\right],
\ee
and
\be \label{ehl}
h_l^{\prime}= \frac{a}{a^{\prime}} \left[3h^{\prime\prime} +9 \frac{a^{\prime}}
{a} h^{\prime} +n^2 h \right].
\ee
The constant $C_e$ reflects the remaining coordinate freedom at the $e$ stage. 
The $C_e$ should be chosen in such a way that the 
comoving synchronous coordinate system of the $m$ stage joins smoothly to the
employed (unique) coordinate system at the $e$ stage; we will discuss this
specific choice of $C_e$ later on. 
The constants $B_1, B_2$ are still arbitrary and should be found from 
solutions at the $i$ stage.  

The ``master equation" at the $i$ stage is
\be \label{imast}
\mu_{n}^{\prime\prime} + \mu_{n} \left[n^2 - 
\frac{(a \sqrt{\gamma})^{\prime\prime}}{a \sqrt{\gamma}}\right] = 0   ,   
\ee 
where 
\[
\gamma \equiv 1 +\left(\frac{a}{a^{\prime}}\right)^{\prime} \equiv
-\frac{\dot H}{H^2}.
\]
and the t-time derivative is related with the $\eta$-time derivative
by $c {\rm d}t =a {\rm d} \eta$. For the power-law scale factors
$a(\eta) \propto |\eta|^{1+\beta}$, which we are working with,
the function $\gamma$ becomes a constant,
and it drops out of the Eq. (\ref{imast}). So, the ``master equation"
(\ref{imast}) is exactly the same as equation (\ref{fieldeq}) for
gravitational waves. By quantum-normalizing the initial metric perturbations,
and evolving them through the $i$ stage, we finally find that
\be \label{B12dp}
B_1 \approx -B_2 \equiv B_{d.p.}. 
\ee
It was shown \cite{gri94} that the crucial quantity $B_{d.p.}$ for density 
perturbations is related with the crucial quantity $B_{g.w.}$ for 
gravitational waves, introduced in Eq. (\ref{B}), by the relationship
\be \label{Bdp} 
B_{d.p.} = \sqrt{6} B_{g.w.}.
\ee
In what follows, we will work with $B_{d.p.}$ only and, henceforth, drop 
the subscript $d.p.$.

Combining all the results together, we write down explicitly the exact 
solution at the $e$ stage, including the required choice of $C_e$. 
In doing this, we use the following new notations:
\be \label{ye}
y \equiv \frac {n}{\sqrt{3}} (\eta - \eta_e),
\ee
\be \label{nc}
y_2 \equiv \frac {n}{\sqrt{3}} (\eta_2 - \eta_e)= 
\frac{n}{\sqrt{3} 2 \sqrt{1+z_{eq}}} =  
\frac {n}{n_c}, ~~~~n_c \equiv 2\sqrt{3} \sqrt{1+z_{eq}},
\ee
\[
Y \equiv \frac{1}{2} y_2 \sin y_2 + \cos y_2 .
\]
Then, the exact solution is: 
\be \label{he}
h(\eta) = \frac{A}{y^2} \left[\cos y - Y \right],
\ee
\be \label{hle}
h_l(\eta) = 3A\left[- \frac{\sin y}{y} - \int_y^{y_2} \frac{\cos y}{y} {\rm d}y-
Y \ln{\frac{y}{y_2}} +\frac{1}{3} \frac{\sin y_2}{y_2} +\frac{2}{3} \cos y_2
\right],
\ee
\be \label{dee}
\frac{\delta \epsilon}{\epsilon_0} = -A\left[ \frac{2}{y^2} (\cos y -Y)+
\frac{2}{y} \sin y - \cos y \right],
\ee
\be \label{vje}
\frac{v^j}{c} = -i A \frac{n^j}{n \sqrt{3}} \left[ \frac{2}{y} (\cos y -Y) +
\sin y \right],
\ee
where
\be \label{A}
A \equiv \frac{i nB \sqrt{1+z_{eq}}}{2 \sqrt{3} l_H} .
\ee

One can check that all the participating functions, $h(\eta), 
h^{\prime}(\eta), h_l(\eta), h_l^{\prime}(\eta),
\delta \epsilon / \epsilon_0, v^j$, join continuously with the
solution (\ref{genm}), (\ref{enden}), (\ref{velz}) at the transition 
point $\eta = \eta_2$. This transition fully determines the 
coefficients $C_1$ and $C_2$:
\be \label{C1C2m}
C_1 = - {A \over 2y_2} \sin y_2,~~C_2 = \frac{3A}{5y_2}
\left[(10 - 3y_2^2)\sin y_2 - 10 y_2 \cos y_2\right].
\ee
The oscillatory behavior of $C_1, C_2$, as functions of $n$, is 
analogous to the oscillatory behavior of the gravitational wave 
coefficients (\ref{C1C2}) and has the same
physical origin. The fact that $B_1 \approx -B_2$ demonstrates that each
mode ${\bf n}$ of the metric perturbations, and the associated matter 
perturbations, at the $e$ stage forms a standing wave pattern. In the 
limit of short waves,
$y \gg 1$, one recovers from Eqs. (\ref{dee}), (\ref{vje}) the familiar
solutions for standing sound waves: 
\be \label{dees}
\frac{\delta \epsilon}{\epsilon_0} \approx A \cos y ,
\ee
\be \label{vjes}
\frac{v^j}{c} \approx -i A \frac{n^j}{n \sqrt{3}} \sin y . 
\ee

\subsection{Perturbations at the last scattering surface}

Having found the quantum-normalized exact solution at the $m$ stage,
we are in a position to calculate the metric power spectrum, which  
is defined by Eq. (\ref{power}). Taking into account our
mode functions, the spectrum can be written as  
\be \label{powerm}
h^2(n, \eta) =  
{{\cal C}^2\over 2\pi^2} n^2 \left[ 
\frac{3}{2} \left|h- \frac{1}{3} h_l\right|^2 + \frac{1}{3} |h_l|^2 \right].
\ee
We will calculate this quantity at the last scattering surface $\eta =\eta_E$.
By that time, the second term in the function $h_l(\eta)$ is 
a factor $[(1+z_{dec})/(1+z_{eq})]^{5/2}$ smaller than the first term
(see Eqs. (\ref{genm}) and (\ref{C1C2m})). We neglect this decaying
part of the solution, participating with the coefficient $C_2$. For the 
explicit form of $|C_1|^2$ we use Eq. (\ref{C1C2m}) and Eq. (\ref{A}). 
Then, we obtain      
\be \label{powermE} 
h^2 (n, \eta_E) = {{\cal C}^2\over 2\pi^2} {n^4|B|^2 (1+z_{eq}) \over
48 l_H^2} \left({\sin y_2 \over y_2}\right)^2 {(300 - 20 p^2 y_2^2 +p^4 y_2^4)
\over 200},
\ee
where we have introduced the quantity
\[
p  \equiv {2 \sqrt{3} \sqrt{1+z_{eq}} \over \sqrt{1+ z_{dec}} },
\]
related to a similar quantity, $b$, from the $g.w.$ case, 
by $p = \sqrt{3} b$.  
The spectrum certainly retains its primordial form in the band of long waves 
$n \ll \sqrt{1+z_{dec}}$. Taking into account Eq. (\ref{Bdp}) and the numerical
value of ${\cal C}=\sqrt{24 \pi} l_{Pl}$ for density perturbations, one 
finds that the primordial
spectrum of metric perturbations associated with density perturbations,
Eq. (\ref{powermE}), is a factor $9/16$ lower than its 
gravitational wave counterpart, Eq. (\ref{powgw}). In particular, 
for $\beta = -2$, one finds
\be \label{powermE2} 
h^2 (n, \eta_E) \approx {9 \over \pi} {l_{Pl}^2 \over l_H^2} {(1+ z_i)^4 
\over (1+z_{eq})}, ~~~ n \ll n_{dec}= \sqrt{1+z_{dec}} .
\ee

\begin{figure}[!hbt] 
\caption{The plot depicted by the solid line is that of $f^2(x,p=8)$
versus $x$. The dashed line shows the behavior of 
$M^2(x)\times 10^6$.}
\centerline{}
\centerline{}
\centerline{}
\centerline{}
\centerline{\psfig{file=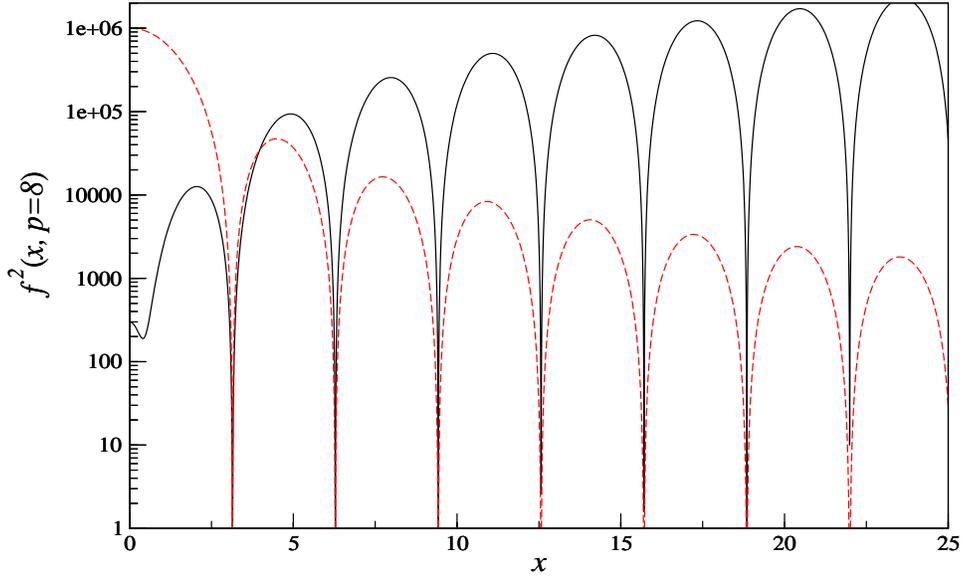,height=3.in,width=5.in}}
\label{fig:fSquare}
\end{figure}

For relatively short waves, $n/n_c \gg 1$, the crucial part of the power 
spectrum (\ref{powermE}) is the modulating (transfer) function
\be \label{transf}
M^2\left({n \over n_c}\right) = \left({\sin y_2 \over y_2}\right)^2 = 
{\sin^2 (n/n_c) \over (n/n_c)^2}.
\ee
The primordial metric spectrum is encoded in the factor $n^4|B|^2$. Whatever 
this spectrum is, the modulating function leaves it intact at large scales,
but bends the spectrum down and introduces oscillations at smaller scales. 
In Fig. \ref{fig:fSquare}, we show 
the metric power spectrum $h^2(n, \eta_E)$, up to numerical
coefficients. (To avoid confusion, we emphasize again that this is the
spectrum of the (squeezed) metric perturbations associated with density
perturbations, and not the gravitational-wave spectrum \cite{LPG02}.) 
Specifically, by a solid line, we plot the function  
$f^{2}(x, p)$, where
\[
f^{2}(x, p) \equiv \left({\sin x \over x}\right)^2 \left[ 300 -20 p^2 x^2 +
p^4 x^4\right],
\]
$x \equiv y_2 = n/n_c$ and, for illustration, we take $p=8$. By a dashed 
line, we plot the function $M^2(x)$, multiplied by the artificial numerical 
factor $10^6$ in order to facilitate the visual comparison of maxima 
and zeros in the two graphs. 

We now turn to the ordinary matter perturbations at the last 
scattering surface. 
The photon-electron-baryon fluid is gravitationally subdominant at $\eta=
\eta_E$. The fluid does not significantly contribute to metric perturbations, 
but it retains its own perturbations. The plasma speed of sound is
given by
\be \label{ss}
{c_l \over c} = {1 \over \sqrt{3(1+R)}},
\ee
where $R = 3 \rho_b/ 4 \rho_{\gamma} \approx 27 \Omega_b h^2$ \cite{wein2},
\cite{wein3}. 
For the popular value $\Omega_b h^2 \approx 0.02$, it means that 
$c_l/c$ decreases from
the nominal value $0.58$ to approximately $0.47$. The plasma standing waves 
(\ref{dees}), (\ref{vjes}), continued to the decoupling era $\eta= \eta_E$, 
take the form:
\be \label{deesE}
\frac{\delta \epsilon_{\gamma}}{\epsilon_{\gamma}} \approx A 
\cos {n \over n_s} ,
\ee
\be \label{vjesE}
\frac{v^j}{c} \approx -i A {c_l \over c} \frac{n^j}{n} \sin {n \over n_s}, 
\ee
where
\be \label{ns}
n_s = {c \over c_l} \sqrt{1 + z_{dec}} \equiv {c \over c_l} n_{dec} . 
\ee
[The velocity $v^j$ is always defined with respect to the unique synchronous
coordinate system, which is comoving with the gravitationally dominant 
pressureless (dark) 
matter.] At the same time, the leading metric perturbation is given by
\be \label{hlE}
h_l(\eta_E) = -{3 \over 5} A {1+z_{eq} \over 1+z_{dec}} {n \over n_c}
\sin {n \over n_c}.
\ee

There are a number of differences between the metric perturbations and
the plasma perturbations at the last scattering surface. 
First, the amplitude
of $h_l(\eta_E)$ is, at least, a factor $3(1+z_{eq})/[5(1+z_{dec})]$ greater 
than the amplitudes of $\delta \epsilon_{\gamma} / \epsilon_{\gamma}$ and
$v^j/c$, near the most interesting scales $n \approx n_c$. After all,
the original motivation for the introduction of a cosmological 
dark matter was precisely this: to avoid
conflicts with $\delta T / T$ observations by allowing the plasma perturbations
at decoupling to be small, but, nevertheless, to be able to develop 
the large scale structure of luminous matter, at the expense of large 
gravitational field perturbations driven by the dark matter. 
So, we have to pay the price for this idea by 
exploring in more detail the consequences of large metric perturbations for
the CMB anisotropies.  

Second, the characteristic frequencies $n_c$ and $n_s$ are different.
Their ratio is
\be \label{ncns}
{n_c \over n_s} = 2 \sqrt{3} {c_l \over c} \sqrt{\frac{1+z_{eq}}{1+z_{dec}}}.
\ee
The $z_{eq}$ is given by 
$1+z_{eq} \approx 4 \times 10^4 \Omega_m h^2$ \cite{wein2}.
For the popular values $\Omega_m = 0.3, h = 0.7$, this amounts to 
$z_{eq} \approx 6 \times 10^3$. So, the ratio $n_c/n_s$ can be a number
close to 4.  
 
Third, although the sound waves before the decoupling are standing waves,
they are still not processed by the quick drop of the sound speed
to zero at the decoupling. This processing will later lead to the 
baryonic matter power spectrum
modulations known as the Sakharov oscillations \cite{zn}. They 
would have taken place even in laboratory conditions, where gravity 
plays no role.
The Sakharov oscillations are important for the formation of 
oscillating features in the luminous matter power spectrum, 
but they are unlikely to be directly responsible for the peaks and dips 
in the observed $C_l$'s. In a broad sense, the periodicity in the
metric power spectrum, related to the transition $\eta = \eta_2$, can 
also be called Sakharov oscillations, but this is not what was
originally meant by the Sakharov oscillations. In short, 
the zeros in the metric power spectrum are 
``frozen" zeros, they are determined by $M^2(n/n_c)$; whereas 
the zeros in the plasma power spectrum, at the times before decoupling,
are still ``moving" zeros; they change their positions at slightly 
different moments of time $\eta = const.$ \footnote{The notion of 
the ``moving zeros" was suggested by J. Peebles in a private 
correspondence, May 1990.}      

Fourth, the wave-number periodicity in the metric power spectrum is 
governed by the sine function, whereas the periodicity in 
Eq. (\ref{deesE}) is governed by the cosine function. Presently, there exists
a tendency to distinguish between the ``acoustic peaks" in the $C_l$  
(supposedly caused by Eq. (\ref{deesE}) and by the ``effective temperature") 
and the ``Doppler peaks" (supposedly caused by the velocity
in Eq. (\ref{vjesE})). The authors of \cite{hudo} emphasize that ``the acoustic
peaks are not ``Doppler peaks"", arguing that the irrotational velocity
cannot produce strong peak structures in the $C_l$ spectrum. They
say that ``the observed peaks must be acoustic peaks" and they give
the ratio of the peak locations: $\ell_1:\ell_2:\ell_3 \sim 1:2:3$ . 
So, the main contenders for the explanation of the peaks 
seem to be the sine function in the metric power spectrum and 
the cosine function in the ``acoustic peaks".

Before proceeding to the discussion of peaks and dips, we need to make
one more comment. It was shown above that the primordial power spectra
of gravitational waves and density perturbations are of the same order
of magnitude, with some small numerical preference for gravitational waves. 
In particular, this is true for the flat spectra 
($\beta = -2$), as demonstrated in Eq. (\ref{powgw2}) and 
Eq. (\ref{powermE2}). Therefore, the lower order CMBR anisotropies (starting
from the quadrupole moment $C_2$) are expected to be of the same order
of magnitude \cite{gri94}. One should be aware that the story is 
dramatically different in inflationary scenario. The ``standard result"
of inflationary scenario \cite{gupi,haw,staro,BST,linde,MFB,LR,LDL}, 
predicts the infinitely
large density perturbations, in the limit of the flat spectrum
(that is, the Harrison-Zel'dovich-Peebles spectrum, with spectral index 
${\rm n}=1$, parameter $\beta = -2$, and the relationship between them being
${\rm n}= 2\beta +5$),
through the set of evaluations: $\delta \rho / \rho \sim h_S \sim
H^2/\dot \varphi \sim V^{3/2}(\varphi)/ V'(\varphi) \sim 1/\sqrt{1 -{\rm n}}$.
By composing the ratio of the gravitational wave amplitude $h_T$ to the
predicted divergent amplitude of the scalar metric perturbations $h_S$
(the so called ``consistency relation": $h_T/h_S \approx \sqrt{1- {\rm n}}$),
inflationary theorists substitute their prediction of arbitrarily large
density perturbations for the claim that it is the amount of gravitational
waves that should be zero, or almost zero, at cosmological scales and,
hence, down to laboratory scales. This claim has led to many years of
mistreatment of a possible $g.w.$ contribution to the CMBR data. It
is only in a few recent papers (for example, \cite{efs}) that the
inflationary ``consistency relation" is not being used when analyzing 
the CMBR and large scale structure observations, with some interesting
conclusions. For the latest statement that the initial spectrum of
gravitational waves is ``constrained to be small compared with 
the initial density spectrum" see the latest article praising inflationary
predictions (for instance, \cite{hudo}). For the critical analysis of
the ``standard inflationary result" see the end of Sec. \ref{sec:dp} 
in \cite{ufn} and references therein.

\subsection{Peaks and dips in the angular power spectrum}  

We will now analyze the zeros and maxima of the metric power spectrum
$f^2(x, p)$ shown in Fig. \ref{fig:fSquare}. The crucial periodic dependence 
is provided by the function $\sin^2(x) \equiv \sin^2(n/n_c)$. We will use this 
function for our analytical evaluation, in full analogy with the case of
gravitational waves. The positions of maxima and zeros are determined 
by the rules:
\be \label{dmax}
{\rm Maxima:}~~ x_k^{max} = {\pi \over 2}(2k+1), ~~~x_k^{max} = x_0^{max}(2k+1),
~~~k = 0,1,2,3,...,
\ee
\be \label{dmin}
{\rm Zeros:}~~ x_k^{min} ={\pi \over 2} 2k,~~~x_k^{min}= x_1^{min}k,
~~~k = 1,2,3,....
\ee
Obviously, the zeros of the function $\sin^2(x)$ are exactly the same as  
the zeros of the full function $f^2(x, p)$. But the positions of
maxima are somewhat different. The difference is significant 
for the zero-th maximum, but it fully disappears for later 
maxima. The locations of the first few maxima, derived from the simple 
analytical formula (\ref{dmax}), are: $x_0^{max} =1.57,~  x_1^{max} =4.71,~  
x_2^{max} =7.85,~  x_3^{max} =11.00$. At the same time, accurate 
positions from the numerical calculation are: $x_0^{max} = 2.05,~ 
x_1^{max} = 4.92,~ x_2^{max} = 7.98,~ x_3^{max} = 11.09$. Thus, formula 
(\ref{dmax}) predicts the positions of the first two maxima somewhat
to the left (smaller $x$) than they should actually appear, but the
positions of zeros and further maxima are described very well.  
In terms of the percentage corrections, the zero-th maximum, derived
from (\ref{dmax}), should be shifted to the right by 30 percent, and the 
first maximum should be shifted to the right by 4 percent.

Accepting $z_{eq} = 6 \times 10^3$, one obtains $n_c = 268$. 
With this $n_c$ and $x_0^{max} = 1.57$, the 
position of the zero-th maximum in the $n$ space would be, 
according to (\ref{dmax}), at $n_0^{max} = 421$.
Positions of all the subsequent features in the power spectrum follow from the
general rules (\ref{dmin}) and (\ref{dmax}). The problem now is to relate
these features in the metric power spectrum with the peaks and dips in the
angular power spectrum $l(l+1)C_l$. Judging grom the previous numerical
experience \cite{dg}, the characteristic features of the metric power 
spectrum are reflected in the $l$-space via a numerical coefficient
$\alpha$ close to $1/2$: $l = \alpha n$. Accepting the provisional
value $\alpha \approx 1/2$, the location of the zero-th peak in the
$l$ space would be near $l_0 = 210$. This is a satisfactory intermediate 
result, but we want to do better. Remembering  that the position of the 
zero-th peak, following from the analytical formula (\ref{dmax}), should 
be shifted to the larger values of $l$, we place our zero-th peak at 
$l_0^{max} = 170$. The 30 percent correction of this number shifts the
zero-th peak to $l_0^{max} = 221$. Of course, we keep an eye on the
actually detected peak in this region. Our aim is to derive the
full structure of peaks and dips in the angular power spectrum $l(l+1)C_l$
from the simple analytical formulas (\ref{dmax}) and (\ref{dmin}),
allowing only for the 30 percent correction to the zero-th peak and the
4 percent correction to the first peak. 
Following this strategy, we formulate our full forecast:
\begin{mathletters}
\label{peaksDips}
\bea \label{peaks}
{\rm Peaks}:&&~ l_0^{max} = 170(221),~ l_1^{max} =510(530),~ l_2^{max} =850,~ 
l_3^{max} =1190,~ l_4^{max} =1530 \ \ ,\\
\label{dips}
&&{\rm Dips}:~ l_1^{min} = 340,~ l_2^{min} = 680,~ l_3^{min} = 1020,~ 
l_4^{min} = 1360,~ l_5^{min} = 1700 \,.
\eea
\end{mathletters}
As a consequence of Eqs. (\ref{dmax}) and (\ref{dmin}), the general rule
for the peak positions is $1:3:5:7...$, for the dip positions: $1:2:3:4...$,
and the dips appear between the peaks at 
\[
l_k^{min} = {1 \over 2} (l_k^{max} + l_{k-1}^{max}).
\] 

Everywhere in this paper, both for gravitational waves and density 
perturbations, we perform calculations under the simplifying assumption
that the Universe is spatially flat. It is obvious, however, that neither 
the generating mechanism itself nor the results, for wavelengths 
comfortably shorter than the putative curvature radius, depend on this 
simplification. The unaccounted factors, such as the possible presence of
a spatial curvature, or a $\Lambda$ term, or a ``quintessence", or 
a ``dark energy", can move the entire structure of peaks and dips, 
but these factors can hardly change the general rules for their relative
positions. 

One should note that what is following from our classification as the
``zero-th gravitational peak", which we place at $l_0^{max} =170$ plus
the correction shifting it to $l_0^{max} = 221$, is often interpreted as 
the ``first Doppler peak" or the ``first acoustic peak". The notion of
the ``zeroth Doppler peak" was introduced and discussed by Weinberg
\cite{wein2}, \cite{wein3}. In general, all three sources - gravitational
field perturbations, intrinsic temperature variations, and velocities - 
contribute to the peak structure. The gravitational field contribution is
represented by the Sachs-Wolfe integral (\ref{sw}), while the two other
sources are represented by Eqs. (\ref{deesE}), (\ref{vjesE}). However,
the raising function $l(l+1)C_l$ would not have turned down without the
modulating function $M^2(n/n_c)$ \cite{dg}, so we focus our attention
on the gravitational contribution.   

The numerical graph of Fig. \ref{fig:fSquare} shows also a little depression 
at $x_{dep} = 0.41$. This depression arises entirely due to the polynomial 
term in $f^2(x, p)$ rather than from the modulating function $M^2(x)$. 
Accepting the same value $n_c = 268$, this feature corresponds to 
$n_{dep} = 110$.
Assigning some significance to this feature, and following the same
logic as before, we have to conclude that this depression in the metric 
power spectrum may be reflected as a small local minimum in the angular power 
spectrum. Applying the numerical factor $\alpha = 1/2$, this minimum
is expected to be seen around $l_{dep} \approx 55$. This may be one of the
areas in the $l$ space to analyse closely in the future experiments,
such as MAP and PLANCK.  

To compare our forecast with observations, we take for the face value
the central positions of peaks and dips reported 
by de Bernardis ${\it et~al.}$ \cite{deb}. We take the liberty of
calling their Peak 1 as our zero-th peak, Peak 2 as the first peak, and
so on. The reported measured positions are as follows:
\begin{mathletters}
\label{peaksDipsObs}
\bea
\label{peaksobs}
{\rm Peaks}:&&~~ l_0^{max} = 213,~~ l_1^{max} =541,~~ l_2^{max} =845 \ \ , \\
\label{dipsobs}
&&{\rm Dips}:~~ l_1^{min} = 416,~~ l_2^{min} = 750 \,.
\eea
\end{mathletters}
Their forecast for the next features is as follows:
\begin{mathletters}
\label{peaksDipsFor}
\bea 
\label{peaksfor}
&&{\rm Peaks}:~~ l_3^{max} = 1139,~~ l_4^{max} =1442 \ \ ,\\ 
\label{dipsfor}
{\rm Dips}:&&~~ l_3^{min} = 1025,~~ l_4^{min} = 1328,~~ l_5^{min} = 1661 \,. 
\eea
\end{mathletters}
Comparing the observed positions (\ref{peaksDipsObs}) with
our formulas (\ref{peaksDips}), we find them in
reasonably good agreement. The peaks and dips appear, at least
roughly, in the right positions.
On the other hand, the periodic function $\cos^2(n/n_s)$, appropriate
for the ``acoustic peaks", implies the reversed  
rules for the dip and peak positions:
\[
{\rm Dips:}~~ l_k^{min} = l_0^{min} (2k+1), ~~k = 0,1,2,3,..;~~
{\rm Peaks:}~~ l_k^{max} = l_1^{max}k , ~~k = 1,2,3,....
\]
So, the structure is supposed to start from the zero-th dip, the dip  
positions are ordered as $1:3:5:7...$, the peak positions are ordered
as $1:2:3:4...$, and the peaks appear between neighboring dips 
at $l_k^{max} = (1/2) (l_k^{min} + l_{k-1}^{min})$. If the first
acoustic peak is at $l \approx 213$, the second one is supposed to be at
$l \approx 426$, almost in the same place where observations indicate
the first dip. Most importantly, there is no observational evidence whatsoever
for the zero-th dip. We do not see how the observed 
structure (\ref{peaksDipsObs}), can be explained by the acoustic peaks.

The forecast (\ref{peaksDips}) on one side, and
the forecast (\ref{peaksDipsFor}) on the other side,
go out of phase at late features. We place our fourth peak inbetween the
positions where de Bernardis ${\it et~ al.}$ \cite{deb} place their 
$l_4^{max}$ and $l_5^{min}$. If these features are not washed out by 
damping \cite{wein2,wein3,hudo} the MAP and PLANCK missions will provide 
the answer. So far, we tentatively conclude
that the structures in the angular power spectrum are caused by
squeezing in the primordial gravitational field perturbations associated
with the density perturbations.

\acknowledgments
We would like to thank P. de Bernardis and P. Mauskopf for a 
helpful discussion on peaks and dips in the angular power spectrum.
Thanks are also due to Bruce Allen, B. S. Sathyaprakash, and Massimo Tinto for 
useful discussions. SB thanks S. Koranda for 
providing the numerical code of Ref. \cite{AK}
and for showing us how to implement it in order to reproduce the results 
given in that reference.
He also thanks A. Dimitropoulos for interesting discussions.
LPG would like to thank K. Glampedakis for help in graphing some
of the equations discussed here. 
The work of SB was funded in part by PPARC grant no. PPA/G/S/1997/00276.
\vspace{2cm}

Note Added in Proofs: 

The latest CBI observations (T.J.Pearson et 
al., astro-ph/0205388) have detected four peaks, at $l \sim 550, 800, 1150, 
1500$, and four dips, at $l \sim 400, 700, 1050, 1400$. These positions are 
in a very good agreement with the theoretical formula (6.35) of the present 
paper. We interpret this data as confirmation of our conclusion that it is 
gravity, and not acoustics, that is responsible for the observed structure.

\end{document}